\documentclass[conference,compsoc]{styles/IEEEtran}
\newcommand{\papertitle}[0]{Issued for Abuse: Measuring the Underground Trade in Code Signing Certificates}

\usepackage{combelow}
\usepackage{amsmath}
\usepackage{amssymb}
\usepackage[normalem]{ulem}  	
\usepackage{soul}				
\usepackage{xcolor}				
\usepackage{url}             			
\usepackage{cite}            			
\usepackage[%
  font=footnotesize, 
  caption=true]{subfig}     			
\usepackage{calc}            			
\usepackage{shadow}          		
\usepackage{fancyvrb}        		
\usepackage{xspace}          		
\usepackage{ifthen}          			
\usepackage{comment}         		
\usepackage{multirow}        		
\usepackage{dcolumn}         		
\usepackage{colortbl}
\usepackage{tabularx}        		
\usepackage{mdwtab}          		
\usepackage{mdwlist}			
\usepackage{ifpdf}           			
\usepackage{rotating}        		
\usepackage{listings}
\usepackage{wrapfig}         		
\usepackage{pdfpages}			
\usepackage{graphicx}        		
\usepackage{fixltx2e}
\usepackage{booktabs, array}          
\usepackage[%
  breaklinks=true,
  pdfkeywords={security, vulnerabilities, cyber attacks, software updates, 
  	telemetry, WINE, empirical studies}]{hyperref}

\makeatletter
\def\@copyrightspace{\relax}
\makeatother

\usepackage[table, subgroups]{svn-multi} 

\svnidlong
{$HeadURL: https://mc2.umiacs.umd.edu/svn/SDSatUMD/papers/2017_code_signing_certs_blackmarkets/main.tex $}
{$LastChangedDate: 2018-06-10 00:04:44 -0400 (Sun, 10 Jun 2018) $}
{$LastChangedRevision: 6305 $}
{$LastChangedBy: tdumitra $}

\def\paperversionmajor{1}          
\def\paperversionminor{\svnrev}    

\def\monthName#1{\ifcase#1\or
  January\or February\or March\or April\or May\or June\or
  July\or August\or September\or October\or November\or December\fi}

\hypersetup{pdftitle={\papertitle}}
\hypersetup{pdfproducer={v. \paperversionmajor.\paperversionminor}}
\ifpdf
\fi

\sethlcolor{lime}

\newcounter{hypothesis}                                     

\newcommand{\topic}[1]{\smallskip \noindent{\bf #1.}}
\newcommand{\question}[1]{\smallskip \noindent{\bf #1?}}
\newcommand{\eat}[1]{} 

\newcommand{\squishlist}{
	\begin{list}{$\bullet$}
		{ \setlength{\itemsep}{4pt}
			\setlength{\parsep}{0pt}
			\setlength{\topsep}{0pt}
			\setlength{\partopsep}{0pt}
			\setlength{\leftmargin}{1.5em}
			\setlength{\labelwidth}{1em}
			\setlength{\labelsep}{0.5em} } }
	
	\newcommand{\squishlisttwo}{
		\begin{list}{$\bullet$}
			{ \setlength{\itemsep}{2pt}
				\setlength{\parsep}{0pt}
				\setlength{\topsep}{0pt}
				\setlength{\partopsep}{0pt}
				\setlength{\leftmargin}{1.0em}
				\setlength{\labelwidth}{1em}
				\setlength{\labelsep}{0.5em} } }
		
		\newcommand{\squishend}{
	\end{list}  }
	
\newcommand{\numBMStudyFinishDate}[0]{2017/08/14\xspace}
\newcommand{\numVTHuntingStartDate}[0]{2017/04/18\xspace}
\newcommand{\numVTHuntingEndDate}[0]{2017/11/24\xspace}
\newcommand{\numGuruCrawlStartDate}[0]{2017/08/25\xspace}
\newcommand{\numGuruCrawlEndDate}[0]{2017/12/07\xspace}

\newcommand{\numGuruSales}[0]{41\xspace}
\newcommand{\numGuruObservationDuration}[0]{104\xspace}
\newcommand{\numGuruCertsPerMonth}[0]{11.8\xspace}
\newcommand{\numGuruTotalRevenue}[0]{\$16,150\xspace}
\newcommand{\prctGuruThawteStockUpdates}[0]{8.7\%\xspace}  
\newcommand{\prctGuruLinkingLikelihoodRandom}[0]{0.0005\%\xspace}  

\newcommand{\numTotalSamplesCount}[0]{2,117,600\xspace}
\newcommand{\numMalwareSamplesCount}[0]{188,421\xspace}
\newcommand{\numCorrectlySignedSamplesCount}[0]{14,221\xspace}  
\newcommand{\numCertificatesCount}[0]{1,163\xspace}
\newcommand{\numPublisherNamesCount}[0]{1,073\xspace}
\newcommand{\numPublisherIdentitiesCount}[0]{788\xspace}  
\newcommand{\numNonSingletonPIsCount}[0]{114\xspace}
\newcommand{\numPercPublishersInNonSingletonPIsCount}[0]{37.19\%\xspace}
\newcommand{\numPublishersInNonSingletonPIsCount}[0]{399\xspace}

\newcommand{\numNonSingletonFamiliesCount}[0]{317\xspace}

\newcommand{\numCertsHavingATimestampedFileCount}[0]{603\xspace}

\newcommand{\numCertsInRussianCluster}[0]{811\xspace}
\newcommand{\numSamplesInRussianCluster}[0]{13337\xspace}
\newcommand{\numPIsInRussianCluster}[0]{477\xspace}

\newcommand{\numNonSingletonFamiliesInRussianCluster}[0]{157\xspace}

\newcommand{\numPercCertsInRussianCluster}[0]{69.73\%\xspace}
\newcommand{\numPercSamplesInRussianCluster}[0]{93.78\%\xspace}
\newcommand{\numPercPIsInRussianCluster}[0]{60.53\%\xspace}

\newcommand{\numPercNonSingletonFamiliesInRussianCluster}[0]{49.53\%\xspace}

\begin{document}

\date{}


\title{\papertitle}

\author{
	\IEEEauthorblockN{Kristi\'{a}n Koz\'{a}k}
	\IEEEauthorblockA{Faculty of Informatics \\ Masaryk University\\
		kkozak@mail.muni.cz}\\ 
	\and
	\IEEEauthorblockN{Bum Jun Kwon}
	\IEEEauthorblockA{University of Maryland\\
		bkwon@umd.edu}
	\and
	\IEEEauthorblockN{Doowon Kim}
	\IEEEauthorblockA{University of Maryland\\
		doowon@cs.umd.edu} 
	\and
	\IEEEauthorblockN{Tudor Dumitra\cb{s}}
	\IEEEauthorblockA{University of Maryland\\
		tdumitra@umiacs.umd.edu}
	
}

\maketitle

\pagenumbering{arabic}

%
%

\begin{abstract}
\noindent 
%
Recent measurements of the Windows code signing certificate ecosystem have highlighted various forms of abuse that allow malware authors to produce malicious code carrying valid digital signatures. 
However, the underground trade that allows miscreants to acquire such certificates 
is not well understood. 
%
%
In this paper, we take a step toward illuminating this trade by investigating the certificate black market from two separate perspectives. 
%
%
First, we identify 4 leading vendors of Authenticode certificates, we document how they conduct business, 
and we estimate their market share.
%
Second, we dig deeper into the demand for code signing certificates by collecting a dataset of recently signed malware and by using it to study the relationships among malware developers, malware families, and certificates.
%
We also 
establish indirect links between these two data sets by inferring that 5 certificates found in our signed malware samples had likely been purchased from one of the black market vendors we observed. 
%
%
%
Using this approach, we document a apparent shift in the methods that malware authors employ to obtain valid digital signatures. 
While prior studies have reported 
that most of the code signing certificates used by malware had been issued to legitimate developers and later compromised, we report that, in 2017, this method is not prevalent anymore. 
Instead, we gather evidence consistent with a stable underground market that represents the leading source of code signing certificates for malware authors.
We also find that the need to bypass platform protections such as Microsoft Defender SmartScreen plays an important role in driving the demand for Authenticode certificates. 
Together, these findings suggest that the trade in certificates issued for abuse represents an emerging segment of the underground economy.

\end{abstract}

%
%

\section{Introduction}
\label{sec:intro}

How can Alice ensure that an executable program developed by Bob has not been tampered with or replaced by malware?
As a solution to this problem, modern computing platforms have adopted 
digital signatures, where the trust derives from a Public Key Infrastructure (PKI).
This infrastructure includes trusted third parties known as Certificate Authorities (CAs), who verify the identity of publishers such as Bob.
Bob, the software publisher, may obtain a code signing certificate from these CAs through a vetting process where his identity is verified.
After acquiring the certificate, Bob can sign his software and embed the certificate that binds the signing keys to Bob's identity.
Alice, the client, can then check the validity of the certificate and verify the signature to ensure that the software really comes from Bob. 

Although a digital signature does not guarantee that the software is safe to execute, 
%
it helps to establish trust in the program. 
In consequence, valid digital signatures can help malware bypass platform protections and anti-virus scanners~\cite{stuxnet-dossier, Kim:2017:codesigningabuse}.
%
%
A well-known example is the Stuxnet worm, which was digitally signed using keys stolen from two Taiwanese semiconductor companies~\cite{stuxnet-dossier}. 
Prior measurement studies have reported many cases of compromised or abused certificates that were embedded in malware~\cite{stuxnet-dossier,goodin_2015,flame_malware,duqu2015, 
	Kim:2017:codesigningabuse} 
or potentially unwanted programs (PUPs)~\cite{Kotzias:2015:CPA:2810103.2813665,Alrawi:2016:CDT:2872518.2888610,
	Wood2010,DBLP:conf/ndss/KwonSDD17}.
While publishers of PUPs (e.g. adware) may obtain code-signing certificates legitimately from CAs~\cite{kotzias2016measuring,DBLP:conf/ndss/KwonSDD17,ThomasCRPPDSTTC16}, malware authors typically aim to conceal their identities when signing their samples. 
As this prevents them from undergoing the CAs' identity verification process, this raises the question \emph{how do malware authors acquire code-signing certificates?}

A longitudinal study of malware signed between 2003--2014~\cite{Kim:2017:codesigningabuse} reported that 
most of the certificates involved had been issued to legitimate software publishers and later compromised, as in Stuxnet's case.
Recent anecdotal evidence suggests that code signing certificates are also traded on underground markets~\cite{GOVrat, CaaS, osborne_2017}. 
This trade allows malware creators to purchase 
a code signing certificate with a fresh publisher identity 
and to use it to sign malware. 

In this paper, we present the first in-depth analysis of this underground trade, considering the whole ecosystem of vendors, malware developers, and certificate issuers, and investigating the vendor's market share and the factors that drive the demand in the market.%
\footnote{A preliminary version of this paper appeared as~\cite{Kozak18:CodeSigningBlackmarket:arxiv}.
In a parallel study~\cite{xrecordedfutureCodesigning}, Recorded Future also monitored 4 certificate vendors but did not analyze in depth the links between these vendors and the signed malware that appeared in the wild during the same period.}
Having inspected 28 forums, 6 link directory websites, 4 general marketplaces and dozens of websites trading black market goods, 
we identify 4 leading vendors of code signing certificates.
The overall activity of these vendors 
has been increasing in the first half of 2017,
with one of the vendors setting up his own e-shop in August 2017.
We regularly collected stock information and analyzed the sales of this e-shop. During the \numGuruObservationDuration days of our observation period, the e-shop obtained a new certificate every two days, on average, and collected \numGuruTotalRevenue in revenue for selling these certificates.
A new code signing certificate generally trades for a few hundred dollars.
Extended Validation (EV) code signing certificates
are also offered for a few thousand dollars each.

To further investigate the impact of this underground trade on signed malware,
we collect \numCorrectlySignedSamplesCount signed malware samples using the VirusTotal Hunting API~\cite{VT-Hunting} and extract \numCertificatesCount certificates from these samples. 
By analyzing the subset of samples that also carry a trusted timestamp, 
we find that around 45\% of the abused certificates are used
to sign malware within the first month after they were issued.
This pattern of abuse is consistent with a black market that frequently obtains new certificates and makes them available for general consumption---as observed in our stock measurements for the code signing e-shop---but would be difficult to sustain by relying on signing keys stolen from legitimate publishers. 
To further corroborate this connection, 
we utilize several properties of the certificates to infer that 5 certificates found in our signed malware samples had likely been purchased from the e-shop during our observation period.


We analyze the relationships among the actors in this underground market by building a graph that connects publisher identities with certificates and signed malware families.
We identify a large strongly connected component that contains most of the samples and mainly Russian publisher identities, with other components being generally small and well separated. 
The strong connectivity within the large component suggests a degree of cooperation among the various malware developer teams and confidence in the black market. 
This cluster also exhibits the highest rate of burning new certificates, 
by using around 60\% of them to sign malware within the first month after issuance.


The emergence of this underground trade points to a growth in the demand for code signing certificates from malware authors.
The requests for certificates on underground forums and the marketing messages of the four certificate vendors suggest that a key factor driving this demand is the need to bypass Microsoft Defender SmartScreen~\cite{SmartScreen}, a certificate reputation system built into Windows 10. 
In turn, this demand has created a market opportunity for specialized vendors who can obtain new certificates regularly and sell them for immediate use.


In summary, this paper makes four contributions:

\begin{itemize}
    \item We conduct an exploratory analysis of the underground trade in code signing certificates. We illuminate its business model, marketplaces, and pattern of transactions.
    \item We report evidence consistent with a shift in 
	how malware is signed, 
	as the demand for new certificates leads to a growing prevalence of certificates issued for abuse (rather than compromised certificates, as reported in the prior work).
    \item We perform a graph-based analysis of the attributes of signed malware, and we infer new relationships among malware developers.
   \item We use indirect evidence to infer that 5 certificates from digitally signed malware found in the wild had likely been purchased from the underground market.
\end{itemize}

We utilize our findings to draw lessons about the role of the certificate black market in the ecosystem of digitally signed malware, and we discuss concrete proposals for facilitating the verification of publisher identities. 
We release our data set of certificates extracted from signed malware samples at \url{https: //signedmalware.org}.

%
%

\section{Problem Statement}
\label{sec:problem}

\subsection{Code Signing Overview}
\label{sec:code-signing-overview}
Microsoft Authenticode~\cite{microsoft2008} is a standard for digitally signing Windows portable executables (PE). 
The digital signatures allow client platforms to identify the publisher who developed the software and to ensure the integrity of the signed executables. 
Each signature appended to an executable is accompanied by a code-signing certificate, which binds the signing key to the publisher's real identity. 
These certificates are issued and signed by Certificate Authorities (CAs), who are trusted to verify the identities of the publishers who request code-signing certificates. 
Each certificate includes a specification of the dates of issuance and expiration.
To avoid the challenge of replacing binaries with expired certificates in the field, software publishers often submit their binaries to Time-Stamping Authorities (TSAs) and receive unforgeable timestamps, proving when the binary was signed. 
Signed binaries that include such trusted timestamps are considered valid even after the certificate expires.
Together, these roles and procedures represent the Authenticode public key infrastructure (PKI), which is the basis for establishing trust in Windows executables.




Unlike the better studied Web PKI, the Authenticode PKI is opaque, as compromised certificates cannot be discovered systematically through network scanning and there is no official list of legitimate software publishers. 
This facilitates abuse, allowing miscreants to obtain code signing certificates and to produce valid digital signatures for 
malicious code.
Several mechanisms have been introduced to prevent this abuse.

\subsubsection*{Revocation Requirements}
In 2016, the Certificate Authority Security Council (CASC) adopted a set of minimum requirements for code signing certificates~\cite{CASC16:MinimumRequirementsBlog}. 
According to these guidelines, a CA must promptly investigate and revoke a code signing certificate after being notified that the certificate has been used to sign malware. 
This makes it difficult for malware authors to reuse code signing certificates. 

\eat{
\subsubsection*{Revocation}
When code signing abuse is detected, certificates, as well as signatures, have to be invalidated in order to prevent further harm. This is done through certificate revocation--certificate is invalidated and information about that is posted to Certificate Revocation List (CRL) as well as to OCSP servers. When client verifies a signature, either CRL or OCSP have to be checked in order to determine that the certificate is still valid.
}

\subsubsection*{EV Certificates}
In addition to standard code-signing certificates, CAs may issue Extended Validation (EV) certificates.
To obtain an EV certificate, the publisher has to pass a more stringent vetting process.
%
EV certificates convey a higher degree of trust and
are required when signing critical code (such as Windows drivers)~\cite{Symantec:EVCerts}. 
In consequence, EV certificates are both more valuable and harder to obtain for malware authors.



\subsubsection*{Microsoft Defender SmartScreen}
To combat malware, \emph{Microsoft Defender SmartScreen} \cite{SmartScreen} assigns reputation scores to executables and Authenticode certificates. 
%
%
When an executable is downloaded on a Windows 10 machine, SmartScreen attempts to assess its reputation before it allows the client to launch it. 
The reputation scores take several factors into account. 
For example, EV certificates receive a good reputation initially. 
Conversely, if the application, the URL from where it was downloaded or its code signing certificate appear on a blacklist, SmartScreen assigns a bad reputation. 
SmartScreen reputations can improve over time, for example when an application garners a track record of installations on multiple hosts without raising suspicions.

However, if SmartScreen encounters a previously unknown application, with a valid 
digital signature that is valid but is endorsed by a previously unknown certificate, Windows 10 displays a warning dialog to the user before launching the application. 
This represents a challenge for malware developers who aim to avoid suspicion and to remain stealthy.
Even if they manage to produce a valid signature for their malware, they must also ensure that their Authenticode certificate has accumulated sufficient reputation in SmartScreen to prevent the user warning.

%

\subsection{Underground Certificate Trade}
\label{sec:underground-certificate-trade}

Despite the anti-abuse mechanisms reviewed above, recent measurement studies have systematically uncovered cases of potentially unwanted programs (PUPs)~\cite{Kotzias:2015:CPA:2810103.2813665,Alrawi:2016:CDT:2872518.2888610,kotzias2016measuring
}
and malware~\cite{Kim:2017:codesigningabuse} carrying valid digital signatures.
These studies have shown that miscreants value the ability to sign malicious code and that they are able to control a range of Authenticode certificates.
However, this line of prior work did not shed light on how malware authors obtain these certificates.%
\footnote{While publishers of PUPs (e.g. adware) typically do not conceal their identities~\cite{kotzias2016measuring,DBLP:conf/ndss/KwonSDD17,ThomasCRPPDSTTC16}, which may allow them to acquire certificates directly from CAs, malware authors cannot rely on this approach.}

At the same time, anecdotal evidence~\cite{GOVrat, CaaS, osborne_2017} suggests that code signing certificates are traded on underground markets.
However, this segment of the underground economy is not well understood. 
Prior research has documented the emergence of similar underground trades in cases where a malicious task required specialized skills and was in demand; for example pay-per-install (PPI) services~\cite{caballero2011measuring, ThomasCRPPDSTTC16, xkotzias2017weis} allow malware developers to focus on implementing the malicious functionality and to outsource the malware delivery task. 
In contrast, the economic forces driving the underground trade in Authenticode certificates have not been analyzed yet.

This is, in part, due to challenges in collecting data about the certificate black market. 
In particular, it is not straightforward to locate the marketplaces where these certificates are being traded. 
%
%
While prior reports mentioned an online marketplace for anonymous code signing certificates~\cite{GOVrat}, this marketplace was no longer active at the time of this writing.
Christin collected and released a large dataset by crawling SilkRoad, a general marketplace for black market goods~\cite{christin2013traveling}; however, our analysis of this dataset did not reveal any certificates or other goods related to code signing abuse. 
Moreover, we were not able to locate any such goods on other general marketplaces available on the dark web, as described later on.

\subsection{Research Questions}
 
Our goal in this paper is to conduct an exploratory analysis of the underground trade in code signing certificates. 
As a first step toward illuminating this black market, we aim to gather evidence that would allow us to formulate hypotheses about the forces driving supply and demand and to suggest further experiments.
Specifically, we ask four research questions:

\question{Q1: What is the business model} 
%
There are three possible business models for monetizing abusive certificates: 
(1) selling the certificates themselves, allowing the malware developers to control them and use them to sign as many malware samples as they wish;
(2) providing a signing service, which receives binary submissions from malware developers and signs these binaries for a fee (in this case, the malware developers do not control the certificates); and
(3) incorporating code signing into Pay-per-Install (PPI) campaigns, where the service providers decide which binaries get signed in order to increase their chances of successful installation. 
We aim to determine which model is prominent today's underground market and to characterize the supply, the demand, and the pricing strategy. 
We are also interested in what drives the demand for certificates---what challenges drive the malware developers to this black market---as this reflects the real-world effectiveness of the defenses against code-signing abuse (reviewed in Section~\ref{sec:code-signing-overview}).


\question{Q2: What are the relationships among the actors involved in the trade}
We aim to investigate the connections among vendors and customers and to assess the stability of the certificate black market. 
Evidence of frequently changing business relationships,
vendors selling the same certificate to multiple customers, 
or malware authors hoarding certificates
could indicate that the trust among these actors is low (no honor among thieves).
Malware authors cannot depend on such a black market for producing valid signatures repeatedly.
In contrast, evidence for communities of malware developers who cooperate and share certificates, and who use newly purchased certificates quickly, could point to a high degree of confidence in the market. 
Such evidence would suggest that the current business models are viable and that this black market could play a growing role in the future malware landscape. 
At the same time, understanding these underground relationships can suggest interventions that would effectively disrupt cyber crime operations~\cite{thomas2015framing}.


\question{Q3: Where do the abusive certificates come from} 
%
Certificates can be either stolen from legitimate publishers or purchased directly from the CAs anonymously (e.g. by setting up a shell company or impersonating a legitimate one). 
Stealing a certificate does not require the malware developer to pay a fee, and it may not be difficult to steal from careless publishers who do not protect their signing keys, but this method does not guarantee a reliable supply of certificates in the future. 
%
%
On the other hand, obtaining a certificate from the CA requires the adversary to pay and should be more difficult to do for malware developers, since they would need to pass the vetting process without revealing their identities. 
However, if adversaries are able to set up a reliable process for passing the CAs' vetting processes, this method would likely scale better with the demand for certificates. 
%
Prior work showed that stealing certificates was the prevalent method between 2003--2014~\cite{Kim:2017:codesigningabuse}. 
We aim to collect a newer data set and to analyze how these trends are shifting, to reason about the role of the certificate black market in the production of digitally signed malware.

\question{Q4: Who controls the abusive certificates}
As the certificates themselves are the most valuable goods in the system of code signing abuse, we aim to analyze signed malware in the wild to infer who controls the abusive certificates---whether it is the malware developers themselves or some third parties that provide code signing services to malware developers.
We also aim to determine if this this evidence is consistent with the business models observed in the vendor analysis from Q1.

%
%

\section{Data Collection}
\label{sec:data}
For our exploratory research, we follow an inductive approach by developing and refining our model of the black market based on the data.
We collect data passively by observing the certificate black market from multiple vantage points. 
We do not interact with any of the actors (e.g. we do not purchase any certificates or exchange messages with the vendors), and we do not conduct any experiments designed to influence their behavior---with the exception of responsibly disclosing the abusive certificates to the CAs. 

To answer Q1 and Q2, we start by exploring the black market. We inspect black market websites where goods related to code signing abuse are traded, analyze vendors and their business models, and observe their activities.
To answer Q2, Q3 and Q4, we analyze samples of signed malware and their connections to certificates and publishers. 
Because certificate sales take place in private, and vendors do not advertise information that would uniquely identify their certificates (e.g. the serial numbers), it is challenging to establish a direct link between the certificates traded on the black market and the ones used to sign malware. 
However, by comparing several distinguishing features of certificates from the two datasets, we infer that a subset of the certificates we found in malware had likely been purchased from one of the vendors we monitor. 
This allows us to connect better the supply and demand sides of Q2. 
In this section, we describe 
our data collection and curation. 

\subsection{Black Market Data}
\label{ss:metds-bm}

To address the data collection challenge discussed in Section~\ref{sec:underground-certificate-trade},
we start with a small number of publicly available, well-known websites (e.g. forums and marketplaces), and use these websites to gradually expand our data set. 
From these data sets, we conduct an investigation on the vendors who trade code signing certificates.
We then analyze information about goods in stock over time to estimate the portion of the black market that we have observed. 

Due to the large variety of both structure and anti-crawling protections the black market sites feature, an automatic analysis is rather difficult. 
Hence, we resort to performing most of the inspection manually, using the sites built-in search tools (search queries are issued for the following strings of keywords: ``code signing certificate", ``code signing", ``sign", ``signature" and ``sign certificate" with Russian equivalents of these being search for as well on Russian speaking forums). 
Automated data collection is utilized where appropriate--e.g. for crawling stock information to see stock updates over time.

\topic{Forums and Marketplaces}
We begin with a set of well known Internet forums (such as \texttt{Hackforums}
), link directory sites (mostly on dark web, e.g. \texttt{Torlinks})
and general marketplaces on dark web (such as \texttt{Dream Market}).
%
We then discover additional sites by following relevant links, searching for user or vendor handles, and searching for the keywords related to code signing certificates on Google and other search engines. 
We expand our data set in this fashion until we stop discovering new sites or we arrive at closed forums that we are unable to access.

In total, we inspect 28 forums, 6 link directory websites pointing to other more specialized web pages, 4 general marketplaces, dozens of websites dedicated to various black market goods (such as credit cards or PayPal accounts), and one website specialized in selling anonymous code signing certificates.
While earlier reports by security companies (such as one by \texttt{InfoArmor} \cite{GOVrat}) reported on an e-shop with anonymous code signing certificates named \texttt{certs4you}, this e-shop was no longer accessible during our research. However, one of the black market vendors observed by us on a forum has set up a new e-shop with anonymous certificates, \texttt{Codesigning Guru}, during the period of our research.
The sites that we investigated are listed in Appendix \ref{app:sites-list}. Screenshots from both forum posts and the e-shop can be found in Appendix \ref{app:screenshots}.

\topic{Vendors \& Purchases}
We have observed in total 4 vendors operating across many forums (under the same handle) and one e-shop, claimed to be set up by one of the observed vendors. We gather information about the vendor activity (registration date, post date, last edit date etc.).

Specifically, we focus on stock updates and vouches. To encourage sales, vendors often provide information about the remaining stock, which can be aggregated across forums. Customers, on the other hand, sometimes provide vouches--a claim that they have utilized the services offered by the vendor and that the offer is not a scam--a mechanism often used on the black market where establishing a trusted relationship is difficult.

On the \texttt{Codesigning Guru} e-shop, payments for certificates were handled through the \emph{Selly}\footnote{www.selly.gg}, a platform that facilitates purchases of digital products paid by digital currencies. 
Through Selly widget, information about the count of certificates on the stock was provided. 
Hence, we have created a crawler that collected count of certificates on stock every five minutes from \numGuruCrawlStartDate till \numGuruCrawlEndDate. 
The use of Selly enabled us to crawl the information by accessing only the Selly widget and not loading the e-shop website itself (in this way, our traffic was more difficult to spot as the e-shop owner would have to get the information about our periodic visits from Selly itself).
This method is similar to the basket inference proposed in \cite{xkanichUseinix2011}, except that we infer the vendor's stock of certificates rather than the content of the customers' baskets.

Information about certificate stock was then used to assess the size of the business and attempt to link the observed sold certificates with certificates observed in the second part--the collection of signed malware samples.

\subsection{Signed Malware Data}
\label{sec:data_malware}

\begin{figure*}
\centering
\includegraphics[width=0.9\linewidth]{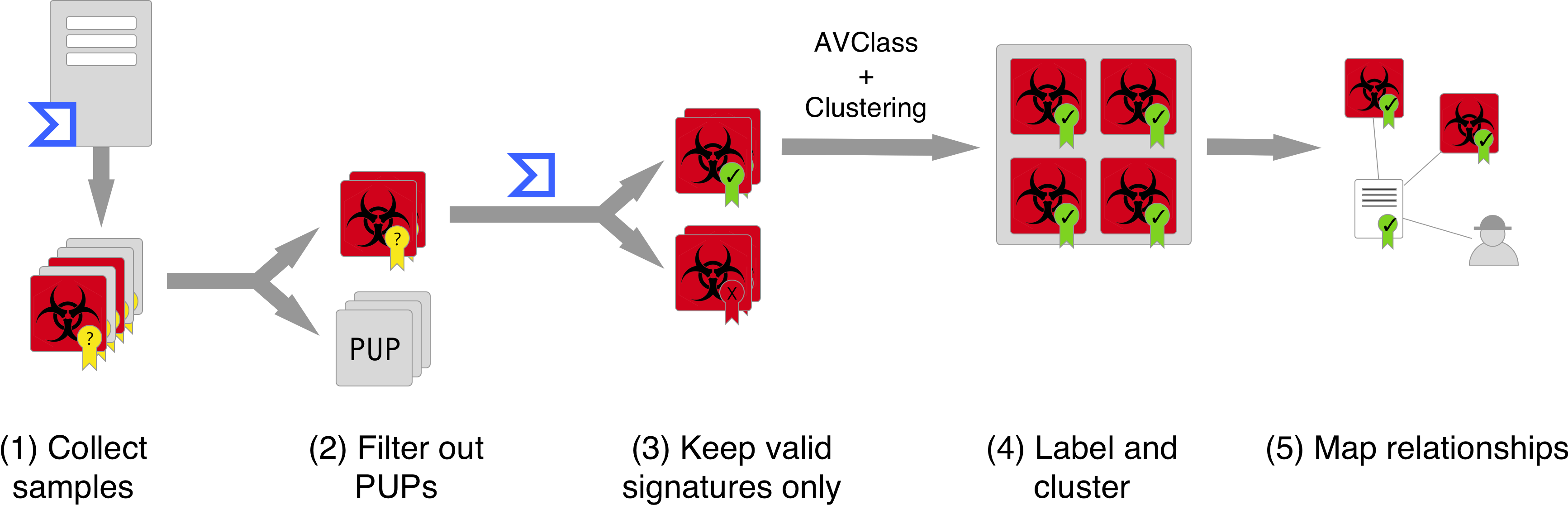}
\caption{Simplified diagram of the signed malware analysis pipeline. Samples were collected on VirusTotal (1), then PUPs were filtered out (2) and only malware with a valid signature was kept (3). Samples were then labeled using AVClass and clustered using Publisher Identities (4). Finally, Malware Map was built (5) to analyze the relationship between the samples, certificates and Publisher Identities.}
\label{f:metds-pipeline}
\end{figure*}

Malware samples were collected over the course of several months. 
Then, PUPs and binaries with unverifiable signatures were filtered out, and only the correctly signed malware was further analyzed.
%


\topic{Data Sources}
Our dataset was collected using the \emph{Hunting} feature on \emph{VirusTotal} \footnote{\url{www.virustotal.com}}--in this way, reports for all files that are submitted to VirusTotal when the hunting is running that satisfy a given condition are collected. In our case, the condition was having more than 10 positive reports and the hunting ran between \numVTHuntingStartDate and \numVTHuntingEndDate.
Since the reports from hunting do not provide detailed information about the signed binary, we re-query VirusTotal with the list of SHA256 hashes caught by hunting using the private API.
The private API returns further information about these hashes including the AV detections and the certificate information--e.g. the issue date, the publisher name and the signing date.

Once our filtering (description of which follows) is applied and we are left with correctly signed malware only, information about certificates are extracted.
%

%

\topic{Filtering for Signed Malware}
To separate PUP from malware, we have used approach introduced in prior work \cite{Kim:2017:codesigningabuse}. First, we have calculated $r_{mal}$ rate as the fraction of positive reports from total AVs that were used to scan the sample. Then, we identified 12 keywords that are often used in AV labels to indicate PUP, such as ``pup" or ``adware". 
We have then calculated $r_{pup}$ rate as the fraction of labels including one of these keywords out of all positive AV reports for given sample.

\begin{figure}
\centering
\includegraphics[width=\linewidth]{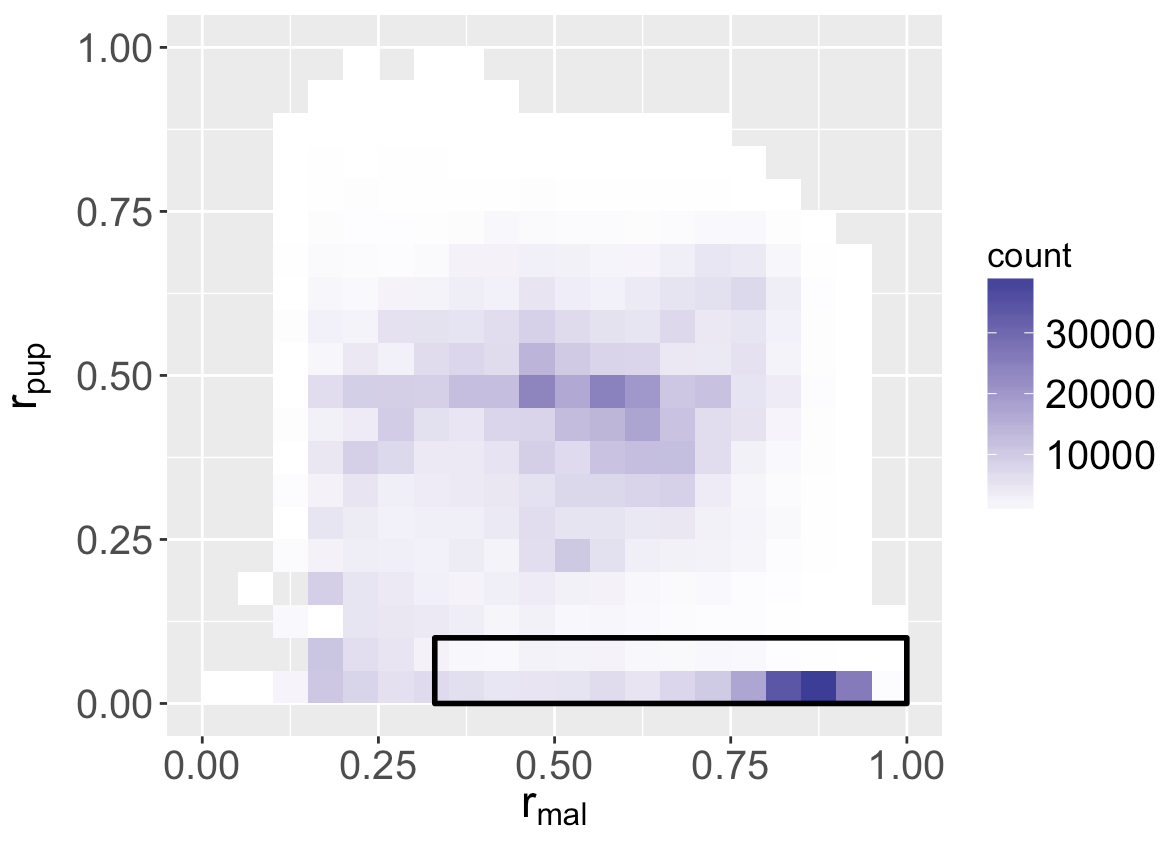}
\caption{Graph showing separation between malware and PUP clusters. White color represents the lowest concentration of samples, violet the largest. Black rectangle shows the threshold values and marks the area of samples that were considered malware}
\label{f:metds-mal_pup_separation}
\end{figure}

While probably not all of the AV labels indicating PUP were caught with our simple 12-keyword approach, Figure \ref{f:metds-mal_pup_separation} shows that there is a clear separation between a cluster of samples that appear to be PUP (having higher $r_{pup}$ and lower $r_{mal}$) and a cluster of samples that appears to be malware with high confidence (very low $r_{pup}$ an high $r_{mal}$). 
Hence, we decided to keep the thresholds used in previous work and recognize samples that have $r_{mal} \geq 0.33$ and $r_{pup} \leq 0.1$ as malware.

Since we want to analyze malware where the developers successfully managed to abuse code signing, we then perform further filtering and keep only samples where the signature was successfully verified by VirusTotal (that is, the ``Verify" flag says ``Signed" in the sigcheck field).
Table \ref{t:metds-counts} provides the summary of the malware samples.

\begin{table}[t!]
\centering
  \begin{tabular}{ @{}lr@{} }
    \toprule
	\textbf{Item} & \textbf{Count} \\ \midrule
	Total samples               &  \numTotalSamplesCount  \\ 
	\quad Malware                     &  \numMalwareSamplesCount   \\ 
	\quad Correctly signed malware    &  \numCorrectlySignedSamplesCount    \\ 
	\midrule
	Samples analyzed		& \numCorrectlySignedSamplesCount \\
	Extracted certificates      &  \numCertificatesCount      \\ 
	Publisher names             &  \numPublisherNamesCount      \\ 
	Publisher Identities (PIs)        &  \numPublisherIdentitiesCount      \\ 
	Non-singleton PIs           &  \numNonSingletonPIsCount       \\ 
	Distinct publisher names in non-singleton PIs & \numPublishersInNonSingletonPIsCount  \\ \bottomrule
  \end{tabular}
  \caption{Summary of counts of samples, certificates and publishers as we proceeded through the data preparation pipeline. Non-singleton PIs are PIs that correspond to at least two distinct publisher names extracted from certificates.}
  \label{t:metds-counts}
\end{table}

\topic{Responsible Disclosure}
We have reported all the certificates we extracted from signed malware to the issuing CAs.

%
%

\section{The Code Signing Black Market}
\label{sec:results-bm}

As explained in Section \ref{ss:metds-bm}, the black market appears to be dominated by 4 vendors (\texttt{A}--\texttt{D}), whose presence ranges from from 2 to 10 forums.
We start our analysis by investigating the business practices of these vendors, aiming to answer our first research question. 
%
%
Table~\ref{tab:bm_vendors} summarizes the periods of activity of these vendors, the apparent sizes of their businesses, and the products they offer. 
%
%
\begin{table*}[t]
    \centering
\begin{tabular}{@{}lrrrrrrlr@{}}
\toprule
\multirow{2}{*}{\textbf{Vendor}}                                                & \multicolumn{3}{c}{\textbf{Activity \& Presence}}                               & \multicolumn{3}{c}{\textbf{Inferred Sales Volume}}              & \multicolumn{2}{c}{\textbf{Products}}                                     \\ \cmidrule(lr){2-4} \cmidrule(l){5-7} \cmidrule(l){8-9}
                                                                                & \textbf{First Joined}       & \textbf{Last Reply}         & \textbf{Forums}     & \textbf{Vouches}   & \textbf{Updates}    & \textbf{E-Shop}      & \textbf{Item: pieces}              & \textbf{Price (\$)}                  \\ \midrule
\multirow{3}{*}{A}                                                              & \multirow{3}{*}{2011-09-07} & \multirow{3}{*}{2015-05-25} & \multirow{3}{*}{3}  & \multirow{3}{*}{0} & \multirow{3}{*}{-}  & \multirow{3}{*}{-}   & Standard: 1                         & 1000                                 \\ 
                                                                                &                             &                             &                     &                    &                     &                      & Standard: 10+                       & 800                                  \\ 
                                                                                &                             &                             &                     &                    &                     &                      & Standard: 15+                       & 700                                  \\ \midrule
\multirow{3}{*}{\begin{tabular}[c]{@{}l@{}}B\\ (Codesigning Guru)\end{tabular}} & \multirow{3}{*}{2016-01-08} & \multirow{3}{*}{2017-08-08} & \multirow{3}{*}{8}  & \multirow{3}{*}{3} & \multirow{3}{*}{7+} & \multirow{3}{*}{\numGuruSales} & Standard (Comodo)                   & 350                                  \\ 
                                                                                &                             &                             &                     &                    &                     &                      & Standard (Thawte)                   & 500                                  \\ 
                                                                                &                             &                             &                     &                    &                     &                      & EV                                 & 2500                                 \\ \midrule
\multirow{2}{*}{C}                                                              & \multirow{2}{*}{2016-09-09} & \multirow{2}{*}{2017-01-27} & \multirow{2}{*}{10} & \multirow{2}{*}{0} & \multirow{2}{*}{-}  & \multirow{2}{*}{-}   & EV (earlier posts)                 & 1600                                 \\ 
                                                                                &                             &                             &                     &                    &                     &                      & EV (current)                       & 3000                                 \\ \midrule
\multirow{6}{*}{D}                                                              & \multirow{6}{*}{2017-03-07} & \multirow{6}{*}{2017-08-13} & \multirow{6}{*}{2}  & \multirow{6}{*}{4} & \multirow{6}{*}{7+} & \multirow{6}{*}{-}   & Standard: 1                         & 400                                  \\ 
                                                                                &                             &                             &                     &                    &                     &                      & Standard: 5                         & 1700                                 \\ 
                                                                                &                             &                             &                     &                    &                     &                      & Standard: 10                        & 3500                                 \\ 
                                                                                &                             &                             &                     &                    &                     &                      & w/ SmartScreeen rep.: 1            & 800                                  \\ 
                                                                                &                             &                             &                     &                    &                     &                      & w/ SmartScreeen rep.: 5            & 3700                                 \\ 
                                                                                &                             &                             &                     &                    &                     &                      & w/ SmartScreeen rep.: 10           & 7000                                 \\ \bottomrule
\end{tabular}
\caption{The leading black market vendors. 
We collected the forum activity on \numBMStudyFinishDate and the e-shop sales from \numGuruCrawlStartDate till \numGuruCrawlEndDate.
Because all the sales take place in private, 
vendors that appear inactive on the forums may still be in business and we cannot observe the sales volume directly. 
We infer the sales volume from vouches, updates of counts of certificates on stock and e-shop stock information crawl (where applicable). }
\label{tab:bm_vendors}
\end{table*}

\subsection{Vendors}
\label{sec:vendors}
%
%
%
\topic{Presence}
Vendor \texttt{D} operated on English speaking forums only (2 forums in total); vendors \texttt{A}, \texttt{C} operated on Russian forums only (10, resp. 3 forums); and vendor \texttt{B}, the most active one, operated on both English and Russian forums (5 English and 3 Russian), though he indicated to be of Russian origin, at one point mentioning he was ``expanding his business [to the English forums]".

Generally, our research started on the more known and publicly accessible forums. However, we have encountered most of the vendors already there and as we continued to progress to forums that are more difficult to access, we still encountered the same vendors again and again. (Eventually, we have reached also forums that we were not able to access due to the closed community.) 
In general, vendors usually start on the more protected forums and then reach out to the more public ones in order to expand their business.

During our observation period, the vendor \texttt{B} has broadened his presence by setting up a new e-shop with anonymous certificates on \texttt{Codesigning Guru} at the beginning of August 2017. Goods and prices generally matched what these vendors were selling on the forums. 
The vendor often promoted the URL of the e-shop on the forums to direct potential customers there.

\topic{Activity}
Generally, the activity of the vendors was surging. While the oldest vendor \texttt{A} appeared not to be very active anymore, with all posts from 2015 with no updates, the two vendors with the largest presence, \texttt{B}, and \texttt{D}, were building it in the last months. The three more recent vendors \texttt{B}, \texttt{C}, \texttt{D} all had half or more of their posts (threads) established after May 2017.

The two most active vendors \texttt{B} and \texttt{D} regularly updated their posts (approx. once or twice per month) and provided replies and stock updates. On major forums, such a thread often contained over 10 public comments of forum users and over 10 replies and updates from the vendor.

\topic{Mechanisms}
Vendors generally start a new thread on a forum, detail their offerings in the first post and then update the post over time with new features, price changes, etc. Other users ask questions and provide vouches (reviews of successful purchases). Vendors answer questions and sometimes provide updates regarding their stock. All purchases take place in private, usually over jabber or messaging service built into the forum platform (which often utilizes jabber as well). Hence, estimating sales and dates of last activity is difficult.

Vendors generally use the same handle (username) across the forums to keep their reputation. The format and content of the post are often exactly the same. For linking vendor identity across forums, we have used vendor contact information, handle, post content and post format--contact information were sometimes missing (relying on the forum's built-in messaging system), then we relied on the other properties.

\topic{Business Model}
At all vendors, the business model was simple: selling anonymous code signing certificates. 
This is surprising, as 
previous reports \cite{GOVrat} 
had reported Signature-as-a-Service providers in the past. 
However, even though we searched for general keywords such as ``signature'' and ``certificate'' (as described in Section \ref{ss:metds-bm}),
we were not able to find evidence of any other business model aside from selling certificates.

All vendors claim that their certificates are freshly issued, that is, that they have not been stolen in any way but obtained directly from a CA. 
Further, all vendors claimed that they sell one certificate into the hands of just one client,
and some even offered free or cheap one-time reissue if the certificate was blacklisted too soon. 
Vendors did not appear to be concerned with revocation, often stating that it usually ``takes ages" until a CA revokes an abused certificate.

%

\subsection{Goods and Deals}

\topic{Goods}
Inventories of all four observed vendors are rather similar. Vendors \texttt{A}, \texttt{D} focus on general code signing certificates, vendor \texttt{C} offers EV certificates only and the most active vendor \texttt{B} (who is behind the \texttt{Codesigning Guru} e-shop) offers both types of certificates.

All vendors claim that certificates are fresh, obtained directly from a CA. For customers, this should be easy to confirm by checking the issue date, as certificates appeared to be sold very soon after being put on the stock by the vendor. Some vendors even claim to obtain the certificate on demand, having the certificate issued once a customer pays half of the price. Interestingly, vendor \texttt{A} even claims that he always has a few publisher identities prepared and the customer can then choose which of these publisher names he wants to have his certificate issued on.


\topic{Certificates with Built SmartScreen Reputation}
Especially on forums that appear more beginner centered, less proficient users often appeared to not have a clear idea about how code signing works, but rather just seeing that SmartScreen blocked their unsigned malware from being launched or produced undesirable warnings. Hence, on these forums, vendors often explain how code signing works in their posts and include keywords like ``bypass SmartScreen" in their post titles indicating that this is what less proficient malware developers search for.

With general (non-EV) code signing certificates, SmartScreen reputation first needs to be built-up before malware can be installed and executed without issuing warnings. 
Vendor \texttt{D} offers certificates with an already built reputation for double the price of a certificate without reputation. Vendor \texttt{B} (\texttt{Codesigning Guru}) does not offer certificates with positive reputation unless a special demand is filed. Moreover, FAQ on \texttt{Codesigning Guru} mentions that approx. 2000-3000 installs of benign files on Windows 10 systems are needed in order to establish sufficient SmartScreen reputation to avoid warnings.

\topic{Prices}
The prices for standard code signing certificates range from \$350 to \$500. 
Vendor \texttt{D} offers certificates from various CAs (citing Comodo, Thawte, DigiCert, Symantec), all for \$400. 
The most active vendor \texttt{B} differentiates pricing by CA: 
a Comodo certificate costs \$350 while Thawte certificates (which are more trusted, according to the vendor) cost \$500 each;
Appendix~\ref{app:ad-post-ex} includes a sample post from this vendor. 
The vendors often offer bulk discounts.

EV certificates are considerably more expensive. Earlier posts by vendor \texttt{C} list a price of \$1600, while more recent posts by the same vendor offer EV certificates for \$3000. Vendor \texttt{B} sells EV certificates for \$2500 (both in the forums and in the \texttt{Codesigning Guru} e-shop). 
The EV certificates come pre-installed on USB tokens, which the vendors then send to the buyers by post. 
As CAs require more rigorous vetting procedures before issuing EV certificates, the supply is more limited than for standard certificates. 
In particular, vendors A and D do not offer EV certificates. 
The posts advertising EV certificates that we collected from vendors B and C did not specify the issuing CA.


\subsection{Sales Volume}

\topic{Forums: Vouches \& Stock Updates}
As all purchases on forums take place in private, it is difficult to estimate sales counts precisely. 
In Table \ref{tab:bm_vendors} we report the number of vouches for each vendor, as vouches are the method for trust on the black market, with users claiming they have used the service and providing a short review of it. 
However, the customers may choose not to submit a vouch even after successful transactions. 
Since we have observed very few vouches (2 for vendor \texttt{B} and 4 for vendor \texttt{D}), estimating sales precisely with this method is not feasible.

A more fruitful resource are stock updates that are sometimes provided by the vendors (even though they should be taken with a grain of salt since they obviously might not be true). For the most active vendor \texttt{B}, we have aggregated in total 17 stock updates across 6 forums. We have for example observed that at one point four code signing certificates were sold in just four days and in another period no certificate sold in 14 days. However, while stock updates might be good to get a very general impression about sales and provide information over short-term periods, estimating long-term sales is challenging.

\topic{E-Shop: Stock Information Crawl}
Our third approach is to analyse
the stock information on the \texttt{Codesigning Guru} e-shop. 
%
%
By using the method described in Section~\ref{ss:metds-bm}, we can infer when certificate sales are completed on this site.

Backed by vendor \texttt{B}, the most active of the four black market vendors, the e-shop offers Comodo and Thawte standard code signing certificates, as well as EV certificates issued by unspecified CAs.
While EV certificates are obtained on demand, the e-shop publishes stock availability for standard certificates.
During our \numGuruObservationDuration-day observation period (from \numGuruCrawlStartDate to \numGuruCrawlEndDate), we have recorded \numGuruSales standard certificate sales.
The average rate of sales for the e-shop was \numGuruCertsPerMonth certificates per month, bringing in a total revenue of \numGuruTotalRevenue (see Table \ref{tab:res-guru-sales}).

\begin{table}[]
\centering
\begin{tabular}{@{}lrrrr@{}}
\toprule
\multirow{2}{*}{\textbf{Certificates}} & \multicolumn{2}{c}{\textbf{Total}}          & \multicolumn{2}{c}{\textbf{Average per month}} \\ \cmidrule(lr){2-3} \cmidrule(l){4-5} 
                                       & \textbf{Items sold} & \textbf{Revenue (\$)} & \textbf{Items sold}   & \textbf{Revenue (\$)}  \\ \midrule
Comodo                                 & 29                  & 10,150                 & 8.36                  & 2,928                   \\
Thawte                                 & 12                  & 6,000                  & 3.46                  & 1,731                   \\
Total                                  & 41                  & 16,150                 & 11.83                 & 4,659                   \\ \bottomrule
\end{tabular}
\caption{Sales volume recorded on Codesigning Guru between \numGuruCrawlStartDate--\numGuruCrawlEndDate.}
\label{tab:res-guru-sales}
\end{table}

The demand appears to be rather high, as the stock availability suggests that certificates are often sold within one day of being put on the stock. 
Additionally, as the vendor claims that the certificates are freshly issued, we can infer the issue dates of the certificates sold on the e-shop; we analyze this information in Section~\ref{sec:res_link}.

\section{Signed Malware Analysis}
\label{sec:results-analysis}

While conducting the study on the code signing black market, we observed that the market is active and the vendors have a potential of scaling the business.
This raises three questions about the demand for certificates: 
(1) how connected the whole ecosystem is---whether an abused certificate could somehow lead us to another abused one, and if there are indications of cooperation among malware developers; 
(2) how prevalent the certificates 
issued for abuse
are among the signed malware and (3) who 
takes control of those certificates.
%

%
We focus on malware since it benefits the most from using anonymous certificates, as code signing helps malware bypass protections~\cite{Kim:2017:codesigningabuse} and anonymous certificates do not reveal the identity of the malware creators.
We collect malware with valid digital signatures, as described in section~\ref{sec:data_malware}. 
To analyze the signed malware ecosystem, we introduce an abstraction, the {\em Signed Malware Map}, that captures the relationships and communities among the malware and the abusive certificates. 
We then map the characteristics and connections in the system in Sections \ref{ss:malmap} and \ref{ss:ecosystem}, examine prevalence of certificates issued for abuse in Section \ref{ss:res-issue-to-abuse}, and investigate who controls the certificates in Section \ref{ss:controllers}.

\subsection{The Signed Malware Map}
\label{ss:malmap}
Here we introduce the {\em Signed Malware Map}, which captures the code signing abuse ecosystem.
The signed malware map is a graph reflecting the relationship among malware families and the certificates they use.
We first introduce the preparation of the entities that the graph consists of.
Then, we describe the construction of the signed malware map.

\topic{Clustering Publishers}
%
%
We observe that multiple certificates from signed malware tend to be issued to what appears to be one publisher company identity with slight variation in the publisher name.
A similar finding was reported for PUP publishers~\cite{Kotzias:2015:CPA:2810103.2813665}.
For the malicious actor who requests these certificates from CAs, this provides the benefit of not having to set up multiple fake identities, while preventing the revocation of multiple certificates belonging to the same malicious publisher.

We tried to apply the publisher clustering technique proposed in prior work~\cite{Kotzias:2015:CPA:2810103.2813665}, which utilizes the normalized edit distance. However, we found out that it was not an effective approach in our case because (1) publishers with different company names with low edit distance (for example companies with short names) ended up in the same cluster even though the publisher identity seemed to be different and (2) most of the certificates did not contain precise street address to help with the clustering.

Hence, we have developed our own approach. Most of the variations in publisher names appeared to be in non-letter characters, such as commas, dashes, quotes and sometimes even backticks. More interestingly, the exact same publisher names have often appeared with different suffices to indicate the company type, such as "Ltd." and "OOO" (Russian version of Ltd.)--even though publisher country stays the same. An example might be the following three publisher names:
\begin{itemize}
    \item Ltd "Vet Faktor"
    \item OOO, Vet - Faktor
    \item LLC "VET FAKTOR"
\end{itemize}
Our technique for clustering the publishers then comprised of (1) removing all company type identification substrings such as "LLC" and "Co.", (2) removing all non-letter characters such as dashes, commas and spaces, (3) converting the string to lowercase and (4) comparing the resulting strings for exact match. 
%
We then examined manually a subset of the samples, to validate the publisher-name clusters produced by our technique. 
We refer to a cluster of publisher names (matched as described above) as a \emph{Publisher Identity (PI)}. 

Over a third (\numPercPublishersInNonSingletonPIsCount) of publisher names extracted from certificates belong to a non-singleton PI (that is, there exists a different publisher name that matches the same Publisher Identity). Count of unique publisher names, PIs, and other statistics are listed in Table \ref{t:metds-counts}. 

\topic{AVClass Labeling}
\label{ss:metds-avclass}
Only clustering of the publishers by itself does not provide enough connections between the certificates and publisher names. Hence, our next step to build analyzable communities was to label the samples with malware family names they are recognized as belonging to--we used AVClass tool \cite{AVClass} to extract the malware family information from AV labels. This step enabled us to build the complete map (network) of malware description of which follows.

For the labeling process, AVClass was left in its default configuration apart from whitelisting the word "confidence" so that AVClass would not consider it a malware family name. This was determined by analysis of the labeled data--after performing the labeling, we have examined outlier families. Family named "confidence" had most samples and certificates, but when we inspected the AV labels that led to individual samples being labeled as "confidence" family, we discovered that this was due to many AVs providing confidence level in the label (e.g. "malicious\_confidence100\%") and AVClass accidentally considering this a family name. Hence, we whitelisted the term "confidence" so that it would not be treated as a malware family name. (Other outliers with many samples and certificates appeared to represent genuine large malware campaigns, such as the labeled Zusy spyware, Loskad trojan, and iCloader and Loadmoney being on the border of PUP and droppers.)

\topic{Building the Signed Malware Map}
To analyze the relationships and communities of PIs and certificates, we have built a graph. There are two types of nodes: (1) nodes representing a malware family and (2) nodes representing a PI. There is an edge between a malware family and a PI if there exists a sample that belongs to this malware family and is signed by a certificate issued to this PI. Moreover, there is an edge between two malware families if there are two samples, each of them belonging to one of the families, that are signed with the \emph{same} certificate. (Hence, the graph is \emph{not} bipartite.) To further simplify the graph, we exclude all singleton malware families--that is, families that contain just one sample, as they do not provide any new connections between publishers and malware families and hence do not contribute significantly to the overview of relationships.

The assumption behind this construction is that if two samples of malware belong to the same malware family, there is a high chance that the same team of malware developers is behind both of the samples. 

We create direct connections between malware families that share a certificate to avoid introducing a third type of nodes representing certificates, which would add further complexity to the graph.
The Signed Malware Map reflects the fact that a connection through a certificate is stronger than through a PI.
In the first case, the same entity controlling the certificate signed malware from both families (whether it's the developer himself or some 3rd party). In the second case, a 3rd party might have just obtained two different certificates for the same PI and shared it with two different malware developer teams (without actually seeing what they are signing with it).

\subsection{Analysis of the Code signing Abuse Ecosystem}
\label{ss:ecosystem}
First, we analyze the structure of the Signed Malware Map, by inspecting the strongly connected components. 
Such components may arise from stable business connections among the participants in the ecosystem.
We have observed one dominant component that contains most of the samples as well as certificates (issued mostly to Russian publishers). 
The remaining nodes are parts of a large number of small components that always contain only a small number of samples and certificates (issued mostly to Chinese publishers). 
While we expected to observe similar distributions distributions of characteristics among malware samples, regardless of which component they are part of, we find interesting differences. 
The
distributions of some characteristics are completely different (e.g. publisher country, issuer) and others are more similar but still feature significant differences (e.g. issue-to-abuse interval).
Because these contrasts may reflect differences in the use of code-signing certificates and in the confidence in the certificate black market, we divide the nodes into these two groups in 
the rest of this section.

\begin{figure}
    \centering
    \includegraphics[width=0.4\linewidth]{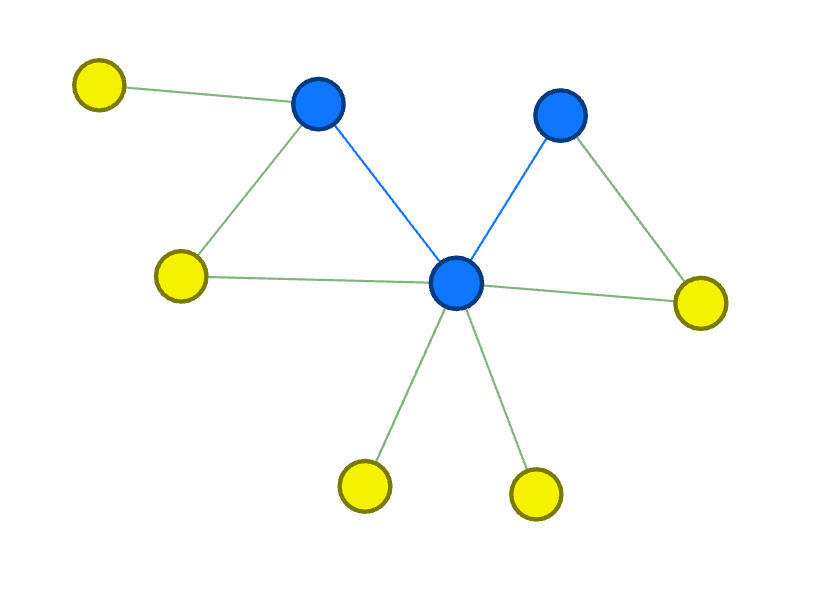}
    \caption{A smaller component consisting of 3 malware families (blue) and 5 PIs (yellow).}
    \label{f:res-an-small-comp}
\end{figure}

\topic{Components}
%
Interestingly, the map consists of one large component (we will refer to it as \emph{Major Component}, or \emph{MC}) which encompasses \numPercCertsInRussianCluster certificates and \numPercPIsInRussianCluster PIs. Other than that, there are 196 small components, none of which contains more than 11 certificates and 9 PIs.
A typical smaller component consisting of 3 malware families and 5 PIs is shown in Figure \ref{f:res-an-small-comp}.

\topic{Connectivity of MC}
Interestingly, if we remove the PI nodes and consider the induced subgraph---that is, we consider only nodes representing malware families and edges between them that represent certificates shared among families---most of the MC remains strongly connected, with only 7 malware families out of total \numNonSingletonFamiliesInRussianCluster getting separated. (Unfortunately, the graph is too big to be included in this paper.)


This strong connectivity suggests that malware developer teams in MC either share certificates or use the same 3rd party service for signing their files. 
As discussed in Section \ref{sec:results-bm}, we did not find evidence of signing services, and the vendors claim that the certificates are freshly issued (a claim we further investigate in Section \ref{ss:res-issue-to-abuse}).
%
In consequence, it appears that malware developer teams themselves control their certificates. 
Then, this type of strong connectivity would indicate strong cooperation among malware developer teams behind samples from the MC.

\begin{table}[t]
    \centering
    \begin{tabular}{ @{}lrrr@{}}
        \toprule
        \textbf{Item} & \textbf{Total} & \textbf{in MC} & \textbf{\% in MC} \\ \midrule
        Samples         & \numCorrectlySignedSamplesCount    & \numSamplesInRussianCluster & \numPercSamplesInRussianCluster \\
        Certificates    & \numCertificatesCount    & \numCertsInRussianCluster        & \numPercCertsInRussianCluster \\
        PIs             & \numPublisherIdentitiesCount    & \numPIsInRussianCluster        & \numPercPIsInRussianCluster \\
        Families        & \numNonSingletonFamiliesCount    & \numNonSingletonFamiliesInRussianCluster        & \numPercNonSingletonFamiliesInRussianCluster \\ \bottomrule
    \end{tabular}
    \caption{Summary of counts of samples, certificates and publishers in the major component. Only non-singleton families (that is, families with more than one sample) are taken into account. MC stands for Major Component.}
    \label{t:anal-mc_stats}
\end{table}

\begin{figure}[t]
    \centering
    \includegraphics[width=\linewidth]{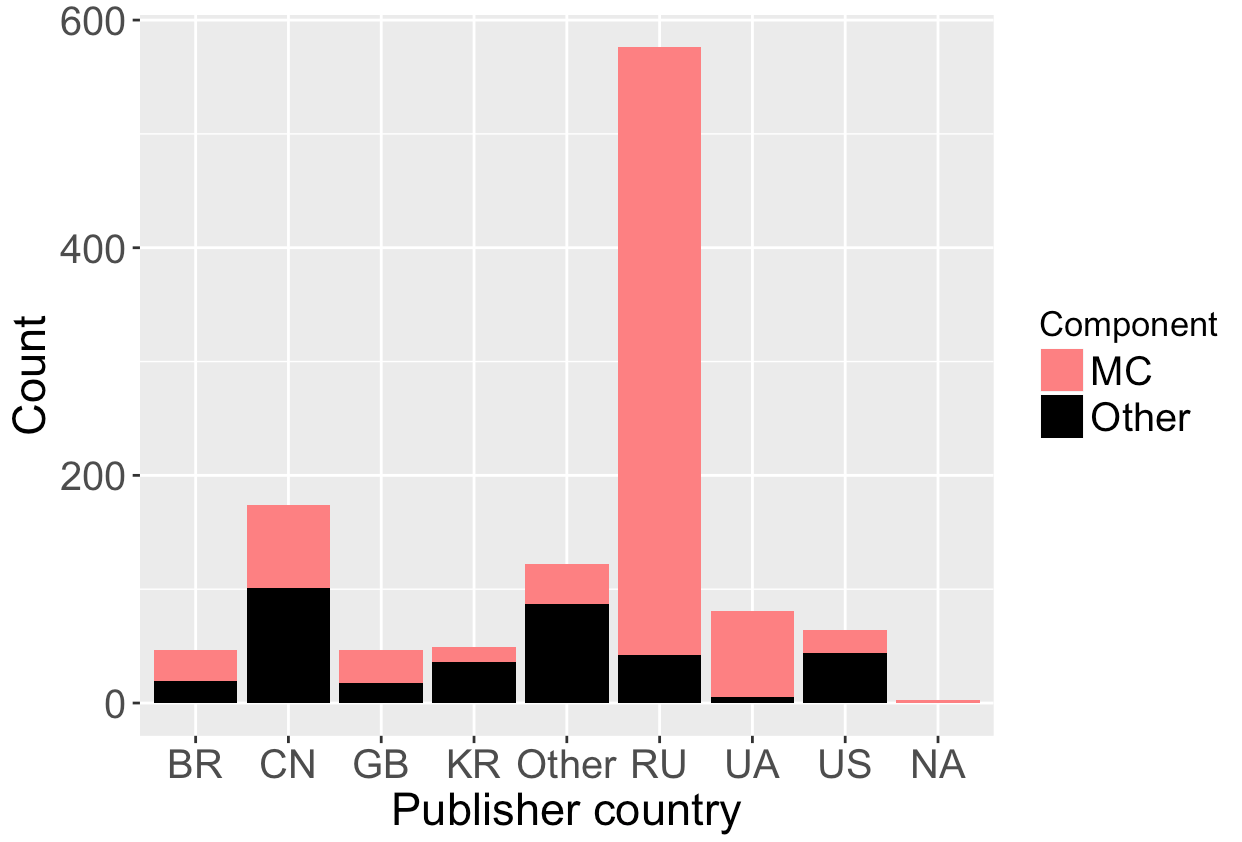}
    \caption{Distribution of publisher countries based on component.}
    \label{f:res-an-mm-countries}
\end{figure}

\topic{Characteristics} 
Table~\ref{t:anal-mc_stats} summarizes the distribution of samples, certificates etc. among the Major Component and other smaller components. 
As we can see, a majority of samples, as well as abused certificates, are in the MC.
At the same time, as shown in Figure \ref{f:res-an-mm-countries}, the majority of certificates in MC are issued to Russian and Ukrainian publishers, while in the smaller components, Chinese certificates are dominant.
Overall, Russian publishers are by far the most prevalent in abused certificates.
This corresponds to the observation that 3 of the 4 black market vendors target Russian speaking audiences (see Section~\ref{sec:vendors}).

\begin{figure}[t]
    \centering
    \includegraphics[width=\linewidth]{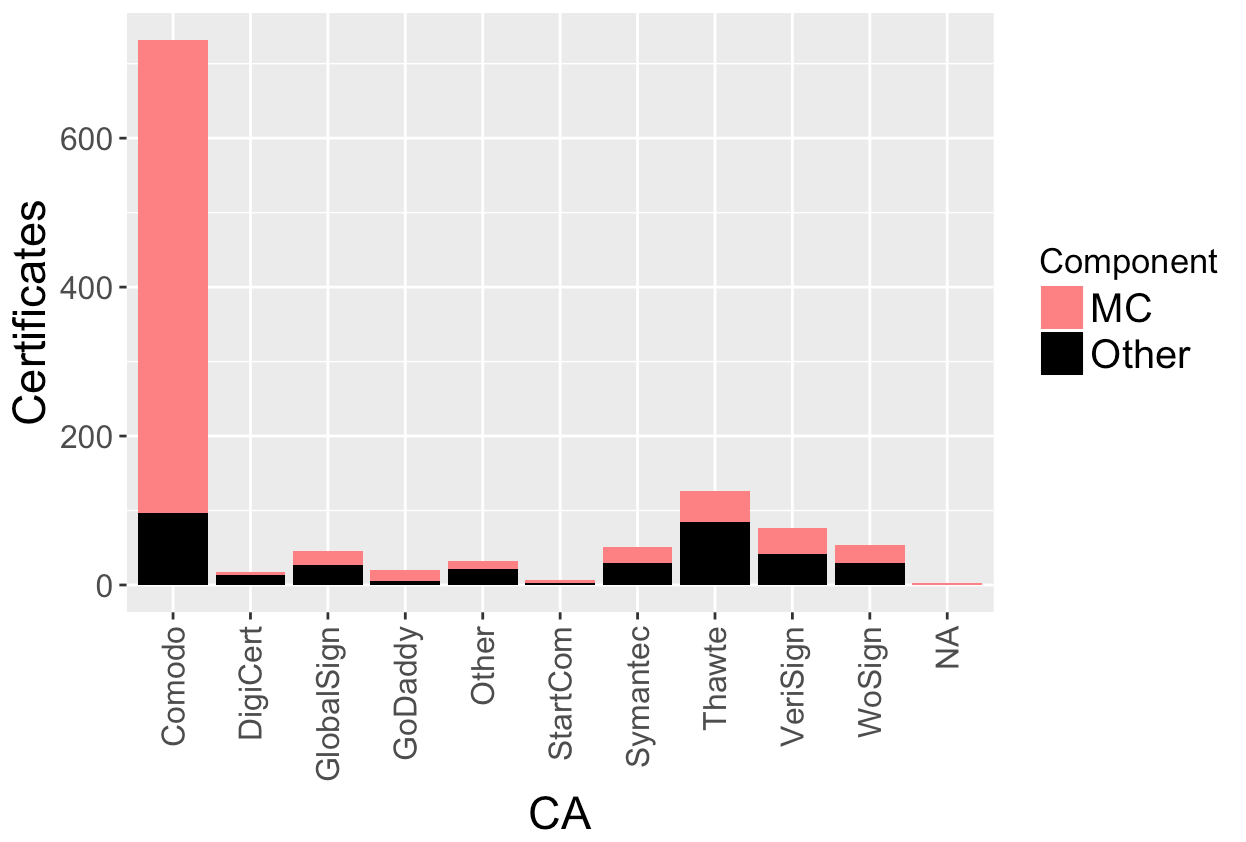}
    \caption{Distribution of CAs based on component.}
    \label{f:res-an-mm-cas}
\end{figure}

Figure \ref{f:res-an-mm-cas} shows the distribution of abused certificates across CAs. 
Most of the certificates from our corpus of signed malware had been issued by Comodo. 
Additionally, Comodo and GoDaddy have more certificates in MC while other CAs have the majority of their abused certificates in the smaller components.

\subsection{The Prevalence of Certificates Issued for Abuse}
\label{ss:res-issue-to-abuse}

\begin{figure*}
    \centering
    \includegraphics[width=\linewidth]{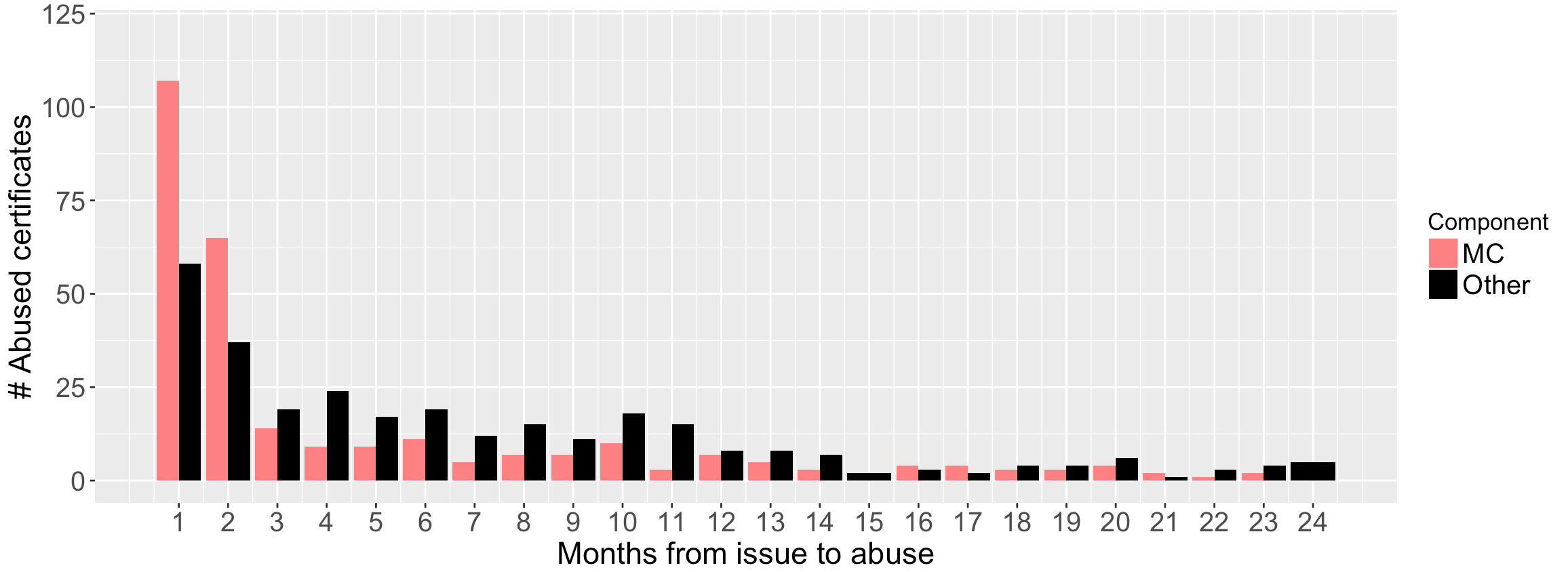}
    \caption{Interval from issue date to the date of first malware being signed and timestamped with a certificate.}
    \label{f:res-an-i2a-hist}
\end{figure*}

We aim to estimate which method of abuse---legitimate but compromised certificates or certificates issued straight for the use of malware authors---accounts for most of the digitally signed malware from 2017.
Prior work \cite{Kim:2017:codesigningabuse} 
identified compromised certificates by searching for certificates used to sign both benign files and malware. 
More recently SmartScreen protections have been built into Windows 10 at operating system level \cite{SmartScreen}, 
driving  
vendors and malware developers to build up SmartScreen reputation by first signing and distributing benign files before using the certificate to sign malware (see Section \ref{sec:results-bm}).
In consequence, the existence of both benign files and malware signed with the same certificate is no longer a strong indication that the certificate was compromised.


We therefore introduce a new technique for estimating how many certificates are issued straight to abusers.
We measure the delay between certificate issue date and the date when the certificate was first abused.
Often, malware developers utilize Trusted Time Stamping to extend the validity of their malware past the certificate's expiration date. 
In total, \numCertsHavingATimestampedFileCount out of the observed \numCertificatesCount abused certificates have been used to sign at least one time-stamped file. 
We utilize these certificates to compute an upper bound for the issue-to-abuse interval, as neither the trusted timestamps nor the issue dates can be tampered with. 
Since the Time Stamp is signed by a Time Stamping Authority, it guarantees that the piece of malware was signed at this date.
For each certificate, we select the file that has the earliest date of Time Stamping and use this date as the date of abuse. 

Hence, we know that certificate was either issued straight to abusers or compromised at some point during this interval. 
For stolen certificates, we expect that they are uniformly likely to be stolen and abused at any point during their lifetime, as suggested by Figure~5 from~\cite{Kim:2017:codesigningabuse}.
However, certificates issued straight to abusers are more likely to be sold and used soon after being issued.
In consequence, a spike of abuse during the first couple of months would be inconsistent with a prevalence of compromised certificates; instead, these certificates would more likely have been traded on the black market.

\topic{Result}
Figure \ref{f:res-an-i2a-hist} shows the distribution of the time-to-abuse (in months), computed as described above and grouped according to whether the certificate is in MC or not. 
Both groups show a tendency for abusing certificates within the first two months after issuance, with more pronounced spikes for the MC.

\begin{figure}
    \centering
    \includegraphics[width=\linewidth]{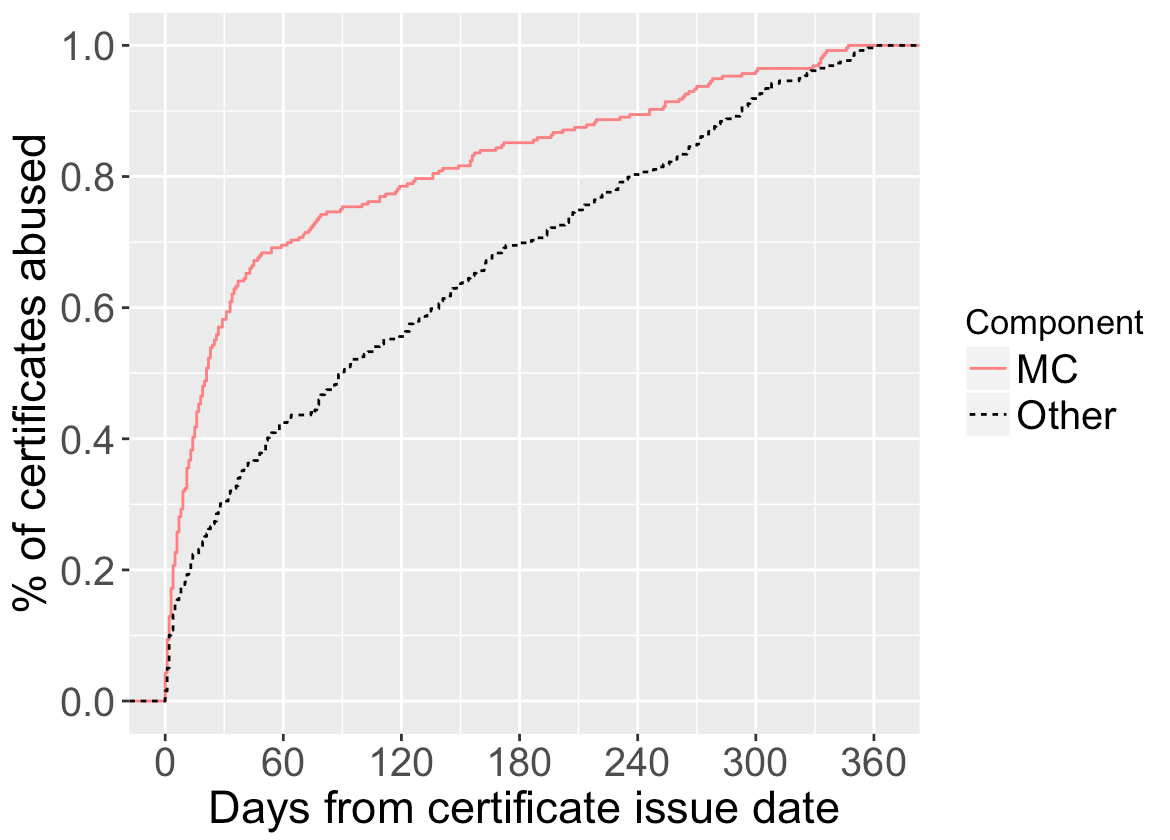}
    \caption{Percentage of certificates abused within specified number of days after being issued.}
    \label{f:res-an-i2a-cum}
\end{figure}

Code signing certificates are usually issued for 1, 2 or 3 years. 
The distribution of the time-to-abuse reflects this fact. 
Between 3--12 months, and between 13--24 months, the time-to-abuse is close to a uniform distribution (this is also true during the 25--36 month interval, excluded from the figure). 
However, there is a sharp spike in the 1st and 2nd month---indicating that a large number of certificates are abused quickly after they are issued. 
Overall, from among all the certificates we extracted from timestamped malware, around 55\% are abused in the first two months (with around 45\% in the first month).
While there are significant spikes for certificates outside of MC, certificates in MC exhibit the sharpest drop after the second month: in MC, more than 10 times more certificates are abused in the first month than on average in months 3-12.
Figure \ref{f:res-an-i2a-cum} shows the same phenomenon as a cumulative distribution; among all certificates that are eventually used to sign malware in MC, around 60\% are abused in the first month, and around 70\% in the first two months, after issuance.

This result corroborates the vendors' claims that they sell fresh certificates, discussed in Section \ref{sec:results-bm}.
Moreover, this timing pattern is unlikely to arise if most certificates used to sign malware are legitimate, but compromised, certificates. 
The certificates abused within the first month were probably obtained with the aim of signing malicious code. 
Additionally, the early spikes in the distribution represent lower bounds for the number of certificates issued straight to abusers, as some malware authors may be saving these certificates for later use.

Outside of MC, more certificates are abused during months 3--12 than inside MC. 
This could correspond to a higher reliance on compromised certificates, but it could also reflect a greater reluctance for burning the certificates quickly, as signing malware may result in the certificate being revoked or losing SmartScreen reputation. 
In contrast, the prevalence of certificates abused within 1--2 months in MC suggests that the malware authors from that community are not worried about burning their code signing certificates. 
This may reflect a confidence in a reliable supply of fresh anonymous certificates, 
perhaps due to tight business connections among underground actors contributing to a higher degree of mutual trust.

In consequence, the distribution of the time-to-abuse is consistent with the hypothesis that most of the certificates found in our corpus of signed and timestamped malware had been issued straight to the black market---at least for a tight-knit underground community that accounts for most of the digitally signed malware.

\subsection{Certificate-Controlling Parties}
\label{ss:controllers}

Our next goal is to infer who controls certificates---whether malware developers themselves control them or if they utilize third-party signing services.
While we did not find any references for signing services during our investigation of the black market from Section~\ref{sec:results-bm}, we aim to corroborate this observation with evidence from the signed malware analysis. 
Specifically, if one certificate signs samples from a large number of families, this would point to a third party who controls the certificates and offers signing services to multiple malware developers.
On the other hand, if a certificate is used to sign only malware from only a small number of families, then would be consistent with a development team that is responsible for these malware families and that controls the certificate.

\begin{figure}
\centering
\includegraphics[width=\linewidth]{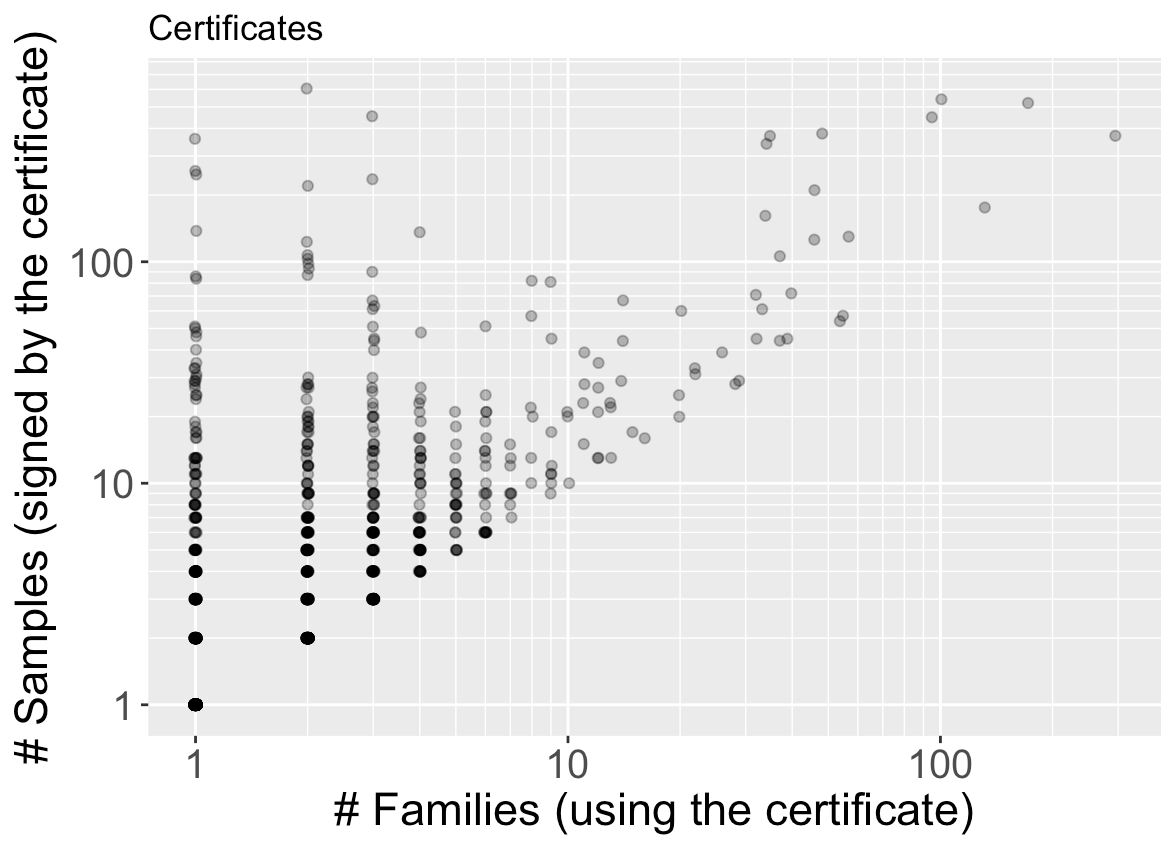}
\caption{Samples and count of families per certificate.}
\label{f:res-an-cvsf-certs}

\vspace{1cm}

\centering
\includegraphics[width=\linewidth]{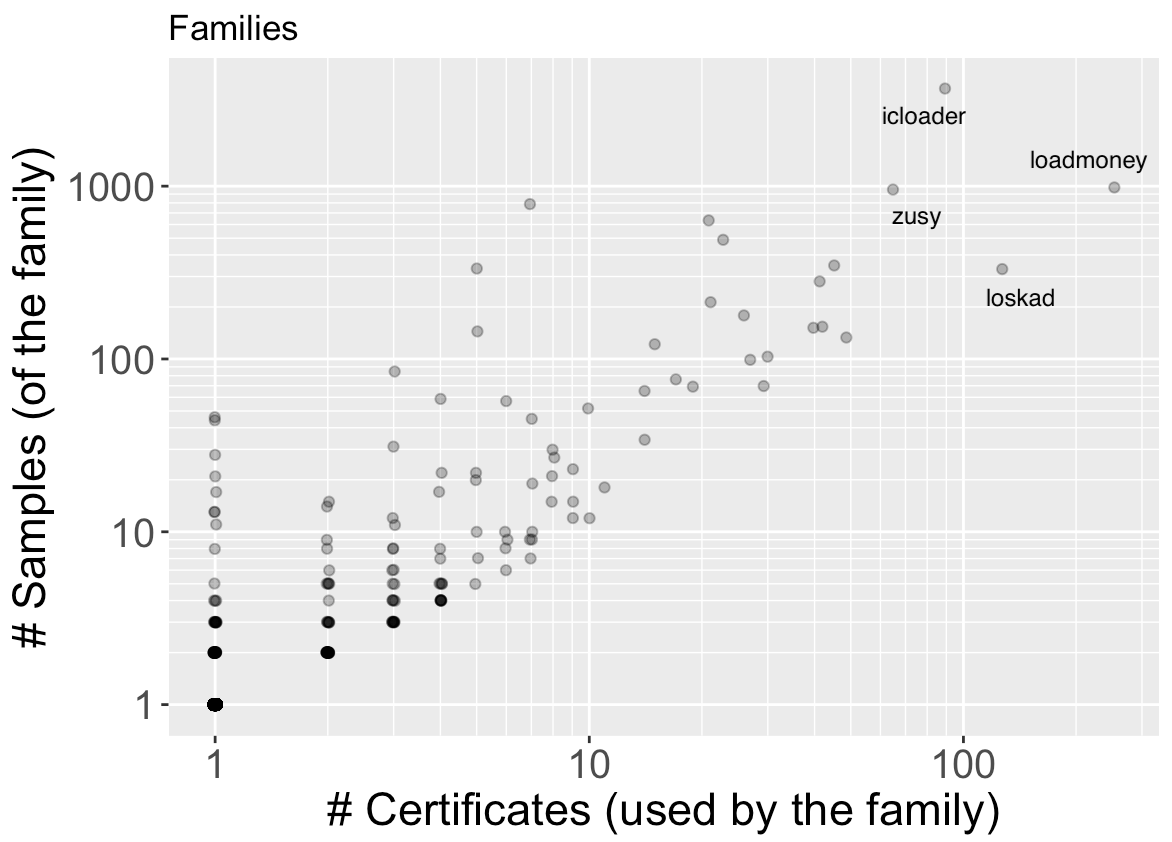}
\caption{Samples and count of certificates per malware family.}
\label{f:res-an-cvsf-fams}
\end{figure}

\topic{Certificate Point of View}
Figure \ref{f:res-an-cvsf-certs} shows that most of the certificates are used to sign fewer than 40 samples among 4 or fewer families---an indication that they are probably in the hands of the specific development team behind these few families. 

Although there are some outliers that signed malware from more than 10 families (with 2 of the certificates having more than one hundred), it seems that signing lot of malware across many families is not the trend. We performed a manual inspection to see what families are signed with these outliers and the families are mostly singletons (families with only one sample). This fact points to two possibilities: either these samples were not recognized by AVs as belonging to one specific family/development team even though they actually do belong to one team, or some signature providing services are used for malware campaigns that comprise of only one or few samples.
However, most of the certificates are used to sign malware across few families, which is consistent with a business model where malware development teams usually acquire code signing certificates for their own use, perhaps by turning to the black market vendors.

\topic{Malware Family Point of View}
Observing the situation from the other side, Figure \ref{f:res-an-cvsf-fams} provides insight on the problem from point of view of malware families. 
Similarly, most families use fewer than 10 samples and fewer than 5 certificates, suggesting that there is usually not much variation in certificates used to sign malware from one family.
This is also consistent with the hypothesis that each development team controls its own  certificates and does not have access to a large inventory of code-signing certificates.
However, we also observe some outliers---malware families with over 500 samples and 100 certificates in some cases.
Figure \ref{f:res-an-cvsf-fams} labels the families that represent the biggest outliers.

\section{Linking Abused and Traded Certificates}
\label{sec:res_link}

To gather additional evidence for the role of the certificate black market in the landscape of digitally signed malware, we aim to determine if some of the certificates from our VT dataset had been purchased from the vendors we monitored.
Unfortunately, it is challenging to identify precisely which certificates originated from the black market.
%
%
%
Establishing a direct link between a traded certificate and a certificate extracted from a malware sample would require matching a unique identifier for the certificate (such as the serial number), but the vendors do not provide such information. 
%
%
Instead, we establish indirect links by matching several certificate properties, which would be unlikely to coincide by chance. 
Specifically, we search our corpus of signed malware samples for certificates that probably had been purchased from the Codesigning Guru e-shop.

\topic{Data Overview}
As explained in Section \ref{ss:metds-bm}, we collected the stock information on the black-market certificate e-shop at 5-minute intervals. 
We made the observation that individual certificates were always added to the stock one-by-one.
Even when 4 certificates were added within 4 hours in one morning, each of them was put on stock separately, with gaps of at least 10 minutes between the stock updates. 

According to the e-shop owner (vendor \texttt{B}), these certificates are fresh, acquired directly from CAs. 
We conjecture that the owner adds certificates to the stock right after obtaining them, with the time required for completing the issuance procedures accounting for the delays between stock updates.
This would imply that the date when a certificate comes on stock usually corresponds to the certificate's issue date. 

Besides the likely issue date, two additional pieces of information can help us narrow down the list of certificates that could have been traded on the e-shop.
For each certificate offered on the e-shop, we also know the issuing CA, as the stock availability is provided separately for Comodo and Thawte certificates (see Table~\ref{tab:res-guru-sales}). 
Furthermore, the e-shop owner claimed on a forum that all certificates sold on Codesigning Guru were issued to British publishers.


\topic{Linking the Certificates}
We compare the dates when individual certificates came on stock to the issue dates of certificates in our VT dataset of signed malware. 
Among all the abused certificates matching the criteria described above, we compute the fraction for which the issue dates coincide with e-shop stock updates. 
This limits the analysis to certificates that are used to sign malware shortly after they are issued; this usage is not uncommon, as discussed in Section~\ref{ss:res-issue-to-abuse}.

We find a strong correlation for Thawte certificates.
%
During our observation period (from \numGuruCrawlStartDate to \numGuruCrawlEndDate), 
11 Thawte certificates came on stock in the e-shop.
However, 2 of these certificates were added during the same 5-min interval, and 2 other certificates were added during the same day, but at different times.
In other words, Thawte stock updates occurred on only 9 days, which represent \prctGuruThawteStockUpdates of the days within our observation period.
During the same \numGuruObservationDuration-day period, 145 certificates from our VT dataset were issued. Out of these, 10 (6.9\%) are from Thawte and 11 (7.6\%) have a British publisher. 
5 certificates used to sign malware from the VT dataset meet all our matching criteria: they were issued by Thawte to a British publisher during our observation period. 
If the CA is equally likely to issue a certificate on any day during this period, the likelihood that the certificate's issue date coincides with an e-shop stock update, by pure chance, is \prctGuruThawteStockUpdates.
In fact, each of 5 matching certificates was issued on a date when a Thawte certificate was added to the e-shop's stock; the likelihood to observe this by chance is \prctGuruLinkingLikelihoodRandom.
%

\topic{Inspecting the Publishers}
As the abused certificates that had likely been purchased from Codesigning Guru represent a substantial portion of the Thawte stock sold there (5 out of 12 certificates, or 42\%), we take a closer look at the publishers named in these certificates to gain additional insights into vendor \textbf{B}'s methods.
%
We search these publishers in the beta version of the British public register of companies\footnote{beta.companieshouse.gov.uk}. 
All the publisher names correspond to young companies, some incorporated around a month before their code signing certificate was issued.
We did not find evidence suggesting that these companies have software development as their primary focus. 
However, 
we were not able to find out the companies' contact information, so we could not confirm their need for using code signing certificates.  
%
%
We hypothesize that these publishers correspond to either (1) shell companies incorporated by a malicious adversary who uses them to request code signing certificates, in an effort to conceal the owner's real identity; or (2) legitimate companies that the adversary is able to impersonate, perhaps by using data mined from the public register. 

%
%

\section{Discussion}
\label{sec:discussion}



%

Our results suggest that it is possible for specialized vendors to set up a reliable process for obtaining code-signing certificates from CAs. 
The underground trade is growing, and at least one segment of the customer base, represented by the MC cluster, demonstrates a degree of confidence in the reliability of the certificate supply. 
While this confidence may be the result of tight business connections among the actors in this cluster, in the future other malware authors may turn to this black market, as long as the vendors continue to provide a reliable supply of code signing certificates. 
This warrants further investigation into the methods the vendors use to pass the CAs' identity verification processes and into the best ways to prevent the abuse. 
Based on our exploratory analysis, we suggest two ways to raise the bar for underground certificate vendors.

\topic{Revoking Matching Publisher Names}
Our research shows that, for over a third (\numPercPublishersInNonSingletonPIsCount) of publisher names we have extracted from the abused certificates, more abused certificates exist that use, in general, the same publisher name with only slight modifications (e.g. added backticks our quotes around the publisher name). Hence, if a CA finds out a certificate has been issued straight to the black market and should be revoked, it would be desirable to also investigate other certificates issued for similarly-named PIs and to revoke them if necessary.

\topic{Standardization of Publisher Name Format}
Since comparing publisher names as we did in our research is not very efficient for large datasets, it would be desirable to standardize the way the publisher name is listed in the certificate (with/without commas, quotes, in the same way as it is listed in a public registry etc.) to make comparing publishers and revoking all certificates issued on (for example) one bogus company identity easier.

%
%

\section{Related work}
\label{sec:related-work}
We discuss related work in the three key areas: (1) measuring anonymous online
marketplaces (i.e., black markets), (2) code signing abuse, and (3) analysis of economic relationships using graphs.

\topic{Black market studies}
Christin~\cite{christin2013traveling} measured {\em Silk Road}, one of the largest anonymous online 
black markets in 2011 and 2012;
the black market was operated as a Tor hidden service.
He found that narcotics were mostly and actively sold and traded in the black market and that
the revenue of this black market was \$1.2 million a month. 
He also showed that the daily sales and the number of the items traded on the black market were increasing during his observation period. 
In October 2013, Silk Road was terminated; then, other anonymous online marketplaces 
(e.g., Silk Road 2, Sheep Marketplace, etc.) started appearing and taking over the anonymous online 
market service of Silk Road.
Sosca et al.~\cite{Soska:2015:MLE:2831143.2831146} analyzed how the Silk Road shut-down 
affected the anonymous online marketplaces by measuring 16 alternative anonymous black markets.
Their observation is in line with Christin's study; most trading items (70\%) were narcotics. 
They also found that only a few vendors were very successful while most vendors earned less than \$10,000
in their entire observation period.

While lots of physical merchandise (e.g. the aforementioned narcotics) are sold on these general black markets such as Silk Road, we were not able to find any offer for code signing certificates on such marketplaces--these appear to be sold usually either on forums or dedicated e-shops.


\topic{Code signing PKI abuse}
%
%
There are a few attempts to examine the code signing PKI abuse and factors. 
First, Sophos~\cite{Wood2010} measured the signed Windows PE files they collected between 2008 and 2010.
They found that the number of signed malicious PE files (e.g., Potentially Unwanted Programs, malware, etc.) was increasing in their study interval.
In line with Sophos' study, Kotzias et al.~\cite{Kotzias:2015:CPA:2810103.2813665} and Alrawi et al.~\cite{Alrawi:2016:CDT:2872518.2888610} also observed that most of the signed malicious PE files they examined were PUPs.
While the abuse factors were not discussed in the previous studies, Kim et al.~\cite{Kim:2017:codesigningabuse} focused solely on signed malware (excluding PUPs) and the abuse factors.
They showed that most signed malware resulted from stolen private keys and also that a lot of signed malware contained an invalid or malformed signature.

Compared to these studies, we focused only on malware that is correctly signed and attempted to document the black market where the certificates came from, which was not previously studied. 
While a majority of samples in the dataset used by Kotzias et al. were PUPs, we excluded PUPs altogether and focused only on malicious signed binaries. 
Compared to Kim et. al, we focused solely on malware where the signatures can be properly verified and used a newer dataset that enabled us to amass much more samples of signed malware and document the current trends--while Kim et. al have shown that in 2014 and earlier, most certificates used to sign malware were obtained by stealing private keys, we document a new trend of obtaining anonymous certificates straight from the CAs that appears to have gained traction during 2016 and 2017.

In a separate study, conducted in parallel with ours, Recorded Future has analyzed the  underground markets trading code signing certificates \cite{xrecordedfutureCodesigning}. While their work focuses on monitoring the certificate vendors and does not collect or analyze a dataset of signed malware, their findings are in line with ours---there are 4 certificate vendors who obtain their certificates directly from the CAs, instead of stealing existing certificates from software developers.
The Recorded Future study also reports an interaction with a vendor, to sign a new malware sample with a recently issued Comodo certificate, in order to confirm the effectiveness of digital signatures in bypassing anti-virus detections. 
In contrast, we collect data passively to avoid influencing the behavior of the underground markets.

\topic{Graph analysis}
Kotzias et al. continued their study of PUPs in 2016 and 2017 by analyzing a network of people and companies behind Spain-based PUP campaigns, using entity graphs and leveraging also information available in code signing certificates used to sign the PUPs \cite{xkotzias2017weis}.
By combining this information with company registers and other sources, they were able to comprehensibly map the PUP network.

The core difference to our work here is that PUP distributors may sign their software using their own identity (and a legitimately obtained certificate),
as PUP programs like adware are not obviously malicious. 
Malware developers, on the other hand, are keen to conceal their identity, hence the need for the underground markets with \emph{anonymous} code signing certificates that we were trying to provide an insight in. 
Therefore, in our analysis, we used the certificates and publisher information in them to create links between malware families, the developer teams behind them, and the black market vendors. 
But we did not attribute these campaigns to real-world people and companies, as the identity of publishers might have been either forged or stolen.

%
%

\section{Conclusions}
\label{sec:conclusions}

We conduct an exploratory analysis of 
the black markets trading code signing certificates. We investigate 4 black market vendors with one of them setting up an e-shop specialized on code signing certificates and selling more than 10 certificates per month with the total of \numGuruTotalRevenue in revenue during our observation period.
Using a dataset of signed malware that we have collected, we support the vendors' claims that the certificates are obtained straight from the CAs by showing that 
around 45\% of all abused certificates are used to sign malware within a month after they are issued.
We further analyze the relationships between the certificates, publishers and malware families to show that individual developer teams appear to be in control of their own certificates. 

Our research provides evidence consistent with a shift in the code-signing abuse ecosystem toward obtaining the certificates straight from the CAs.
Once specialized vendors establish reliable processes for obtaining certificates in this manner, this model scales better with the demand than a model that relies on compromised certificates. 
%
%
We suggest two practical ways to make this abuse more difficult: 
searching for certificates issued to similarly named publishers and revoking them as appropriate,  
and standardizing the format for publisher names.
More importantly our exploratory results explain the existence of a growing black market for code-signing certificates, and they warrant further investigation into the methods the underground vendors use to obtain fresh certificates.



\section*{Acknowledgment}

\noindent
The authors would like to thank the anonymous reviewers for their constructive suggestions, Nicolas Christin for searching the SilkRoad data set for goods related to code signing abuse, and
VirusTotal for allowing us to collect our corpus of signed malware samples. 
This research was partially supported by the National Science Foundation, through award CNS-1564143 and two travel grants.

{



\bibliographystyle{abbrv}


\bibliography{bibliography/security}
}

\appendices

\section{List of Inspected Sites}
\label{app:sites-list}

In total, 1 e-shop specializing in code-signing certificates, 28 forums, 4 general markets and 6 link directory sites were inspected during the black market investigation--all of these are listed below. Dozens of sites specialized on other goods (such as stolen Credit Cards and Paypal accounts) that were mentioned at the listed link directory sites were inspected as well, but they are not listed here since (1) as expected, they did not provide any value or hints of code-signing certificates, (2) this absence of value was not surprising (as opposed to general markets where we expected to find anonymous code signing certificates), (3) their addresses were extracted from the link directory sites listed here and (4) the space in this section is limited.

Note that the Alphabay general market was not inspected since this research was carried out shortly after Alphabay has been taken down by law enforcement. All the URLs are given as they were valid during our research period (especially some of the darknet ones might not be valid anymore).

\subsection{E-Shop with Anonymous Code-Signing Certificates}
\noindent
\texttt{codesigning.guru}

\subsection{Forums}
\noindent
\texttt{hackforums.net}\\
\texttt{antichat.ru}\\
\texttt{0day.su}\\
\texttt{freehacks.ru}\\
\texttt{nulled.to}\\
\texttt{bitcointalk.org}\\
\texttt{sinister.ly}\\
\texttt{hpc.name}\\
\texttt{searchengines.guru}\\
\texttt{binaryvision.co.il}\\
\texttt{bitcointalk.org}\\
\texttt{offensivecommunity.net}\\
\texttt{russianelite.ws}
\texttt{verified.ws} \\ 
\texttt{fsell.bz}\\
\texttt{forum.zloy.bs}\\
\texttt{elitepvpers.com}\\
\texttt{carder.su}\\
\texttt{prologic.su}\\
\texttt{thecc.bz}\\
\texttt{abusewith.us}\\
\texttt{v3rmillion.net}\\
\texttt{Crutop.nu}\\
\texttt{cardx.biz}\\
\texttt{cop.su}\\
\texttt{wordcarding.su}\\
\texttt{psh-world.ru}\\
\texttt{crimezone.org}\\
\texttt{elitepvpers.com/forum}\\

\subsection{General Markets}
\noindent
\texttt{hansamkt2rr6nfg3.onion} (Hansa)\\
\texttt{lchudifyeqm4ldjj.onion} (Dream Market)\\
\texttt{traderouteilbgzt.onion} (Trade Route)\\
\texttt{trdealmgn4uvm42g.onion} (theRealDeal)\\

\subsection{Link Directory Sites}
\noindent
\texttt{deepdotweb.com}\\
\texttt{torlinkbgs6aabns.onion} (Torlinks)\\
\texttt{underdj5ziov3ic7.onion} (UnderDir)\\
\texttt{onionsnjajzkhm5g.onion}\\
\texttt{zgrl6sghf5jh37zz.onion}\\
\texttt{tt3j2x4k5ycaa5zt.onion}\\

\section{Forum and E-Shop Screenshots}
\label{app:screenshots}

\begin{figure}[h]
    \centering
    \includegraphics[width=\linewidth]{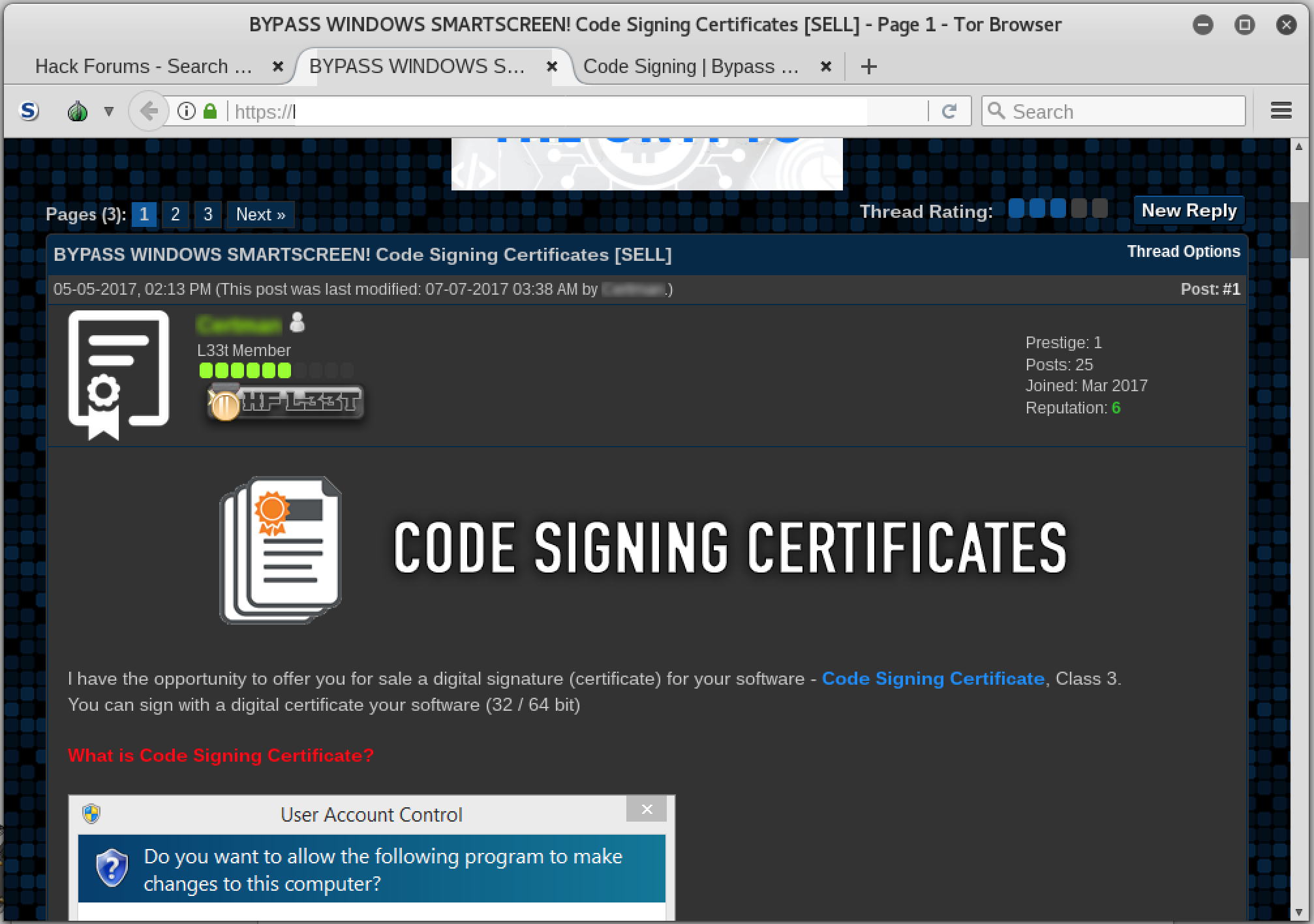}
    \caption{Example post offering code signing certificates on the black market.}
    \label{fig:ap-forum-hf}
\end{figure}

\begin{figure}
    \centering
    \includegraphics[width=\linewidth]{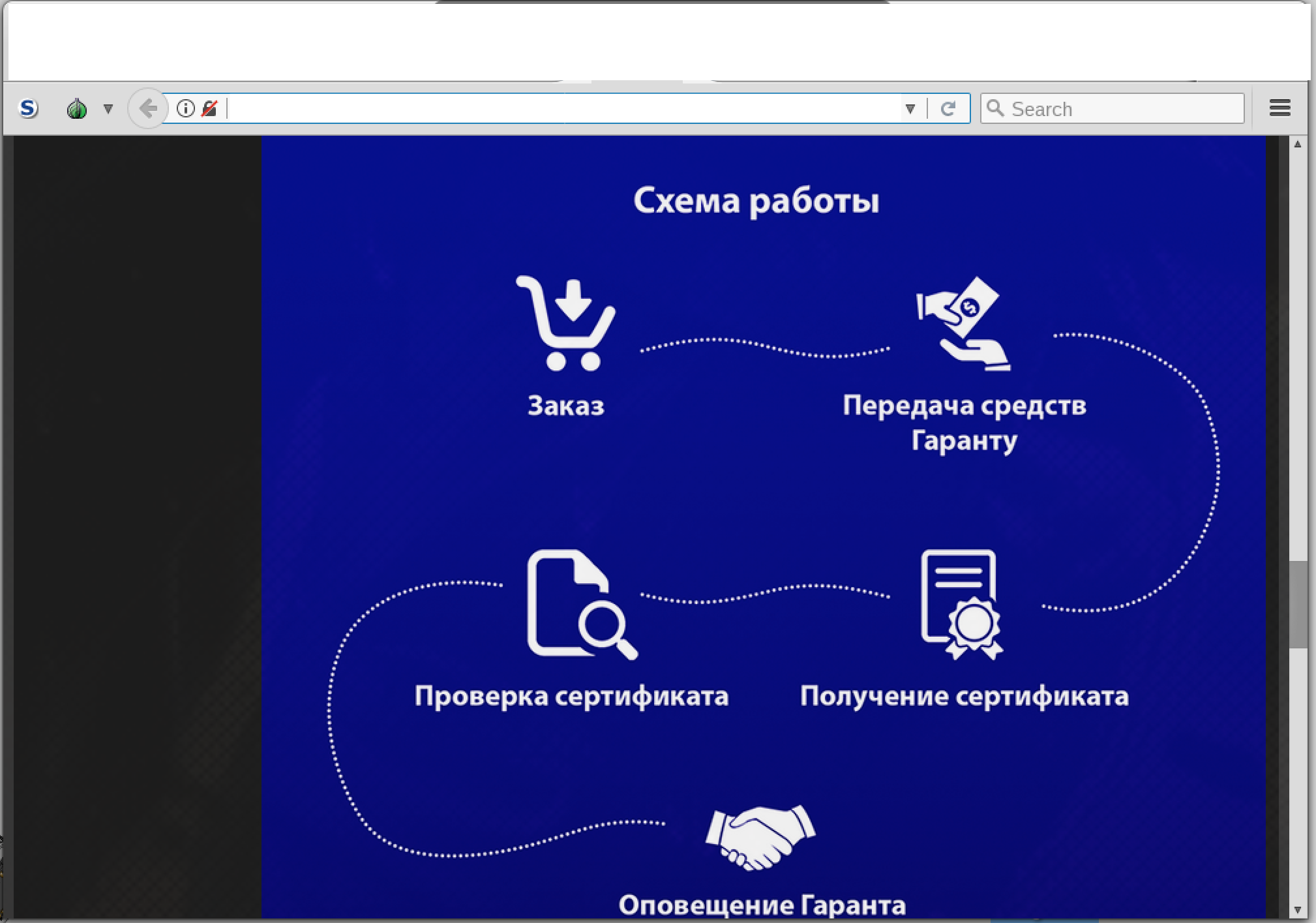}
    \caption{Code signing certificate purchase workflow illustrated in a black market post.}
    \label{fig:ap-forum-workflow}
\end{figure}

Code signing certificates are often advertised on black market forums. Some posts include even rich graphics. An example post is shown in Figure \ref{fig:ap-forum-hf} and Figure \ref{fig:ap-forum-workflow} depicts a diagram of code signing certificate purchase from other black market post.

\begin{figure}
    \centering
    \includegraphics[width=\linewidth]{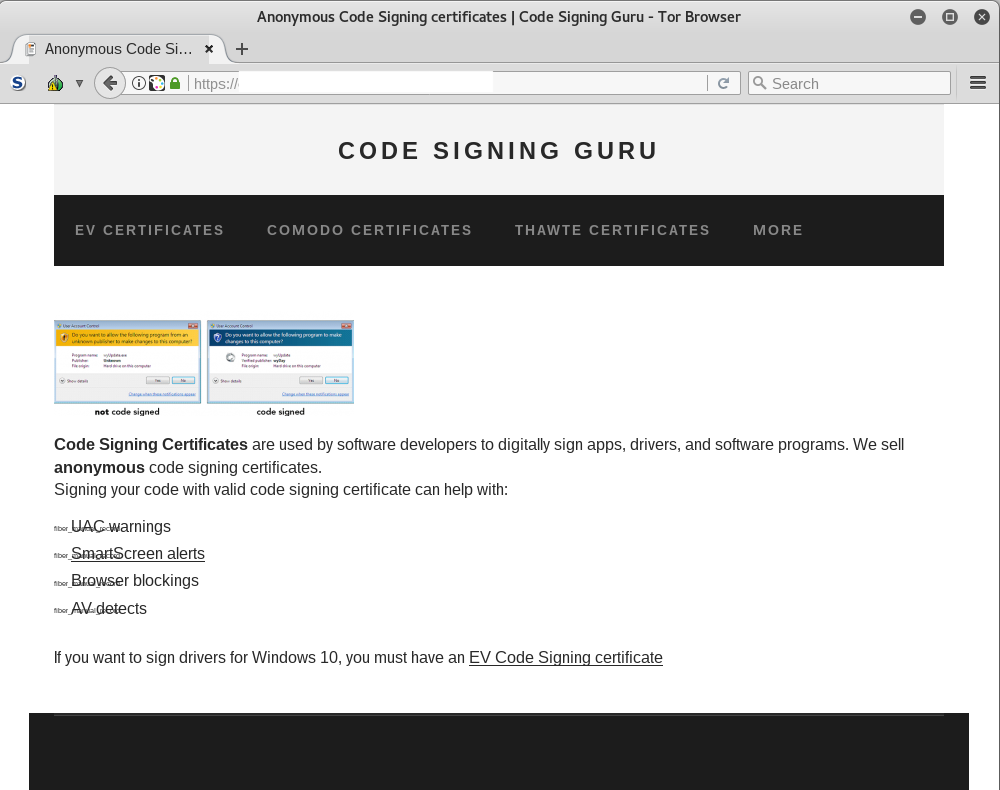}
    \caption{Main page of the codesigning guru e-shop.}
    \label{fig:ap-guru-home}
\end{figure}

\begin{figure}
    \centering
    \includegraphics[width=\linewidth]{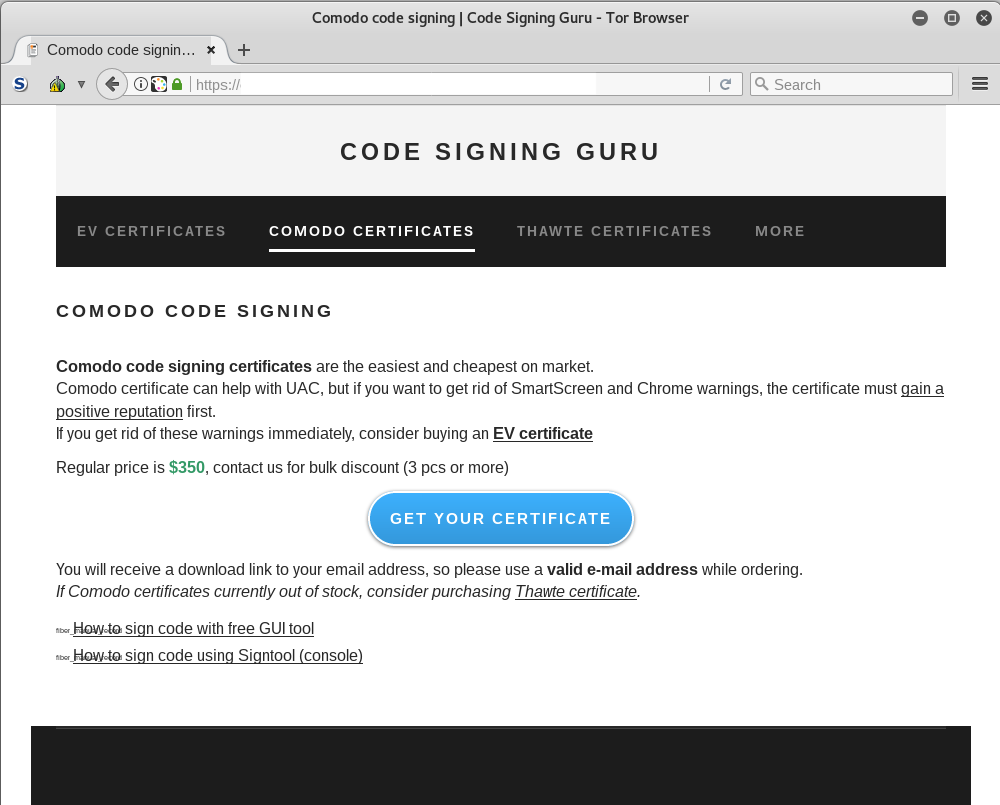}
    \caption{Page for buying Comodo code signing certificates in the codesigning guru e-shop.}
    \label{fig:ap-guru-buy}
\end{figure}

The main page of the codesigning guru e-shop is shown in Figure \ref{fig:ap-guru-home} along with the page where Comodo code signing certificates can be purchased in Figure \ref{fig:ap-guru-buy}.

\newpage
\section{Advertisement Post Example}\label{app:ad-post-ex}

\begin{Verbatim}[frame=single]
I'm selling anonymous code signing cert-
ificates.

Why do I need to sign my files?
- to avoid UAC warnings
- to pass SmartScreen filter* ("Unknown 
publisher")
- to pass some AVs which are blocking 
any unsigned executables
- to make your macros and VBA objects 
more trusted

What files can I sign?
- 32 and 64-bit applications (.exe, 
.cab, .dll, .ocx, .msi, .xpi and .xap)
- Drivers (EV certificates only)
- Java applications
- Apple applications
- VBA objects, scripts and macros for 
Microsoft Office .doc, .xls, and .ppt 
files

Certificate types
- regular code signing certificate 
(class 3): you'll get the archive 
with PFX and a password.
Regular certs should gain a reputation 
before they pass SmartScreen filter 
(contact me for details)

- EV Code Signing cert: you'll get a 
USB token with pre-installed certificate 
via mail or courier.
EV certificates are the only ones that 
you can use for signing drivers for 
Win10.
Also they have a positive SmartScreen 
reputation out-of-the-box, so no SS 
warnings will appear.

What you get
- Unique and never used before 
Code Signing certificate from trusted 
Certificate Authority, valid for 1 year.
- Free code signing tool with GUI 
(if needed)
- All certificates are issued for 
real companies

Disclaimer
- I DO NOT resell certificates 
from other sellers.
- I guarantee that your certificate is 
unique.
- I DO NOT offer one-time signing.

Pricing
- Comodo (most common): $350
- Thawte (more trusted than Comodo): 
$600
- EV code signing certs: $3000
- Free exchange: if your cert becomes 
blacklisted by AVs, you can get one 
re-issue for free.
- Escrow accepted
- All payments by Bitcoin only.

Current stock:
- Comodo: 2
- Thawte: 0 (pre-order and get it in 
3-5 days)
- EV Code Signing: 0 (pre-order and 
get it in 2-3 weeks)
Contacts
- Jabber: <e-mail address removed> 
(use OTR please)
- PM me 
\end{Verbatim}

\end{document}